**Applying and testing a novel method to estimate animal density from motion-triggered cameras**

Marcus Becker [*1]

David J. Huggard [2]

Melanie Dickie [1,3]

Camille Warbington [1]

Jim Schieck [1]

Emily Herdman [4]

Robert Serrouya [1]

Stan Boutin [5]

* Corresponding author:

mabecker@ualberta.ca

(780) 492-6322

[1] *Alberta Biodiversity Monitoring Institute, University of Alberta, Edmonton, AB, T6G 2E9, Canada*

[2] *Apophenia Consulting, North Vancouver, BC, V7L 2E7, Canada*

[3] *Department of Biology, University of British Columbia, Kelowna, BC, V1V 1V7, Canada*

[4] *Innotech Alberta, Edmonton, AB, T6N 1E4, Canada*

[5] *Department of Biological Sciences, University of Alberta, Edmonton, AB, T6G 2E9, Canada*

**Abstract**

Estimating animal abundance and density are fundamental goals of many wildlife monitoring programs. Camera trapping has become an increasingly popular tool to achieve these monitoring goals due to recent advances in modeling approaches and the capacity to simultaneously collect data on multiple species. However, estimating the density of unmarked populations continues to be problematic due to the difficulty in implementing complex modeling approaches, low precision of estimates, and absence of rigor in testing of model assumptions and their influence on results. Here, we describe a novel approach that uses still image camera traps to estimate animal density without the need for individual identification, based on the Time spent In Front of the Camera (TIFC). Using results from a large-scale multi-species monitoring program with nearly 3,000 cameras deployed over six years in Alberta, Canada, we provide a reproducible methodology to estimate parameters and we test key assumptions of the TIFC model. We compare moose (*Alces alces*) density estimates from aerial surveys and TIFC, including incorporating correction factors for known TIFC assumption violations. The resulting corrected TIFC density estimates are comparable to aerial density estimates. We discuss the limitations of the TIFC method and areas needing further investigation, including the need for long-term monitoring of assumption violations and the number of cameras necessary to provide precise estimates. Despite the challenges of assumption violations and high measurement error, cameras and the TIFC method can provide useful alternative or complementary animal density estimates for multi-species monitoring when compared to traditional monitoring methods.

## Introduction

Reliable estimates of animal abundance and changes in abundance over time and space are a fundamental component of many ecological studies and monitoring programs (Mills, 2007; Mace et al., 2008). To estimate the abundance of mid- and large-sized mammals, researchers have developed multiple sampling techniques, including those that rely on direct visual detections (e.g. aerial surveys) or evidence of past presence (e.g. snow tracking, scat counts) (Wilson and Delahay, 2001). Because they are easy to implement and relatively less invasive, remote cameras have emerged as a popular alternative or complementary tool for wildlife monitoring (Trolle et al., 2008; Burton et al., 2015; Amin et al., 2016; Steenweg et al., 2017).

As remote camera use has increased, so too have the analytical approaches to estimating abundance with camera data (eg., Royle et al., 2009; Gilbert et al., 2021). Early applications focused on species with marked populations and applied well-established capture-recapture techniques to estimate population size (Karanth and Nichols, 1998). However, individual identification is impossible or impractical for many applications and species (Rayan et al., 2012; Palmer et al., 2018), warranting alternative approaches. Methods proposed to allow estimation of abundance for unmarked animal populations include space-to-event (Moeller et al., 2018), distance sampling (Howe et al., 2017) and spatial count models (Chandler and Royle, 2013; Gilbert et al., 2021). Rowcliffe et al., (2008) proposed the Random Encounter Model (REM), which estimates density as a function of the encounter rate (measured by the number of images collected per unit time), animal movement speed, and the area of the camera's detection viewshed. Accurate measurement of animal speed is a key challenge hindering widespread adoption of the REM because it requires either telemetry data or intensive observations of behavior (Rovero and Marshall, 2009; Zero et al., 2013; Caravaggi et al., 2016; Rowcliffe et al., 2016; Gilbert et al., 2021). To address this limitation, Nakashima et al., (2018) proposed a modification to the REM that incorporates ‘staying time’ – the time spent in the camera detection zone – as a proxy for movement speed: the Random Encounter and Staying Time (REST) model. Because staying time is inversely proportional to movement speeds, it can be substituted as a model parameter. Staying time can be measured from images or video collected by remote cameras (Nakashima et al., 2018). This modification improves the accessibility of the REM approach to practitioners, but does not change sampling design considerations. Importantly, both models require representative (or random) placement of cameras relative to animal movement in order to obtain unbiased estimates of abundance (Nakashima et al., 2018).

While REST eliminates the need to measure animal movement speed, its implementation is mathematically challenging (Bessone et al., 2020) and requires video, which limits its usefulness for most camera trap programs. Two recent studies make use of data collected by cameras to measure animal staying time, but do not require parameterization of separate encounter rates and staying times (Laurent et al., 2020; Warbington and Boyce, 2020). Warbington and Boyce (2020) called this method Time In Front of the Camera (TIFC), a term that we adopt here. TIFC treats camera image data as quadrat samples and can be used to estimate the density of unmarked animal populations directly from still image camera traps. This method has also been tested in principle by Garland et al (2020) using human volunteers.

In this paper, we describe the underlying theory of the TIFC method, outline how it can be operationalized, and explore potential biases that could arise under typical field conditions. First, we explain the TIFC method as a modification of traditional quadrat sampling, and identify the information required to apply TIFC density estimation using motion-activated cameras that collect still images. We also discuss the use of lure to increase detection rates, and how this technique can be integrated into a monitoring program. Next, we outline key assumptions of the TIFC model: random or representative location of the cameras relative to animal movement, no influence of the cameras on animal movements, and reliable detection of animals in at least part of the camera field-of-view. We illustrate how these three assumptions pose different limitations to our results, and suggest potential ways to measure them and correct for violations. As an example, we focus on the results for moose (*Alces alces*), but highlight other species that show different patterns in these violations. Additional results for all species commonly detected by our cameras are included in Appendices. Finally, we compare moose density estimates based on TIFC and aerial distance sampling surveys conducted in Alberta, Canada, in order to understand how TIFC performs relative to an established monitoring approach.

## Methods

### *Using cameras as quadrats*

Density, which is the number of individuals per unit area, is often estimated by ecologists using fixed-area plots ("quadrats") (Manly, 2014). For instance, if a 100m$^2$ quadrat contains a single tree, then the density is 1 tree per 100m$^2$. If the quadrat can be considered a representative sample of a larger area, the density can then be extrapolated to apply to the greater area, e.g., 10,000 trees per km$^2$ in this example. Camera traps can conceptually be considered a quadrat by counting the number of individuals within the camera's field-of-view (i.e. detection zone). If a camera has a field-of-view of 100m$^2$ and there is one animal continually present in the field-of-view, then density is one animal per 100m$^2$, or 10,000 animals per km$^2$ in the broader region that the camera represents. However, unlike trees, animals move. To address this trait, we take advantage of a unique feature of camera traps relative to typical quadrat sampling protocols:

continuous sampling conducted over long periods of time. We can therefore calculate the number of animals in the field-of-view as the average number of animals over time, including all the time when no animal is present. If one animal is present for 1/10,000 of the total camera operation time, the density is 1/10,000 animals per $100m^2$, or 1 animal per $km^2$. Two animals present for the same duration would give 2 animals per $km^2$. Thus,

$$D = \frac{\sum(N \cdot T_F)}{A_F \cdot T_O} \tag{1}$$

where $N$ is the number of individuals, $T_F$ is the time in the camera field-of-view, $A_F$ is the area of the field-of-view, and $T_O$ is the total camera operating time. The units are animal-time per area-time, which reduces to animals per area. To determine density, the important variable is total animal-time in the field-of-view, whether that comes from one long visit by one individual, several shorter visits from one individual, or several shorter visits by different individuals. For a given density of animals, this simple measure is independent of both home range size and movement rates (Garland et al., 2020). If movement rates were twice as fast, then an individual would spend half as much time in the field-of-view, but would pass by a camera twice as often. Similarly, if home ranges were twice as large, an individual would be in a camera's field-of-view half as often, as there are twice as many other places for it to be. However, twice as many animals would have home ranges overlapping that camera.

Unlike traditional quadrat sampling, cameras have additional complexity because a camera's field-of-view (the area being sampled, $A_F$) is not fixed: the probability of an animal triggering the camera decreases with distance (Rowcliffe et al., 2011). In addition, converting the discrete images taken by a motion-detection camera into total time in field-of-view requires additional analysis, including adjusting for the possibility that an animal leaves the field-of-view between images and returns. In the next sections, we apply distance sampling methods to estimate effective detection distances and, by extension, the area surveyed by a camera, which can vary between species, habitat types, and time of year (Buckland et al., 2015; Apps and McNutt, 2018; Hofmeester et al 2019). We then address the time in front of the camera components of the TIFC model by converting discrete images into time in front of the camera. This procedure, and the analyses that follow, are presented as a workflow in a public repository that can be accessed here: *https://github.com/mabecker89/tifc-method*.

*Data collection*

We applied TIFC using remote camera data collected throughout the province of Alberta, Canada. We used a systematic-random sampling design, based on a grid pattern of point locations spaced 20-km apart plus a random offset up to 5.5 km to maintain confidentiality of exact locations. At each of these sites, we placed four Reconyx PC900 Professional Hyperfire cameras at each corner of a square with 600-m long sides (ABMI, 2019). We used a scent lure (O'Gorman Long Distance Call) at two of the four camera deployments at each site. Cameras were mounted to either a tree or a stake, with the base of the camera unit 1m from the ground. We placed a 1 m tall brightly colored pole 5 m in front of each camera to facilitate effective detection distance analysis (discussed below). We aimed the camera at the base of the pole. Data were collected from 2014 to 2020, totaling 2,990 cameras across 799 sites. Cameras were deployed between November and February, and retrieved later that year in July or August. The median number of operational days was 149 per camera deployment (mean of 162, SD of 59).

*Components of the model*

*Effective area of the camera field-of-view ($A_F$).* Effective detection distance (EDD) is the distance from the camera that would give the same number of detections if all animals up to that distance were perfectly detected and none were detected farther away (Buckland, 1987). We used the pole placed 5 m from the camera to divide animals in images into two distance bins: < 5 m from the camera and > 5 m from the camera. We excluded images of animals too close to the pole to classify or animals actively investigating the pole or camera. We only used images from unlured cameras to estimate the EDD. From these data,

$$EDD\ (m) = \frac{5}{sqrt(p_{<5m})} \qquad (2)$$

where $p_{<5m}$ is the proportion of images that contain animals between the pole and the camera. The calculation assumes that we detected all animals that occurred within 5m of the camera (tested below in Assumption 3). The area of the camera's field-of-view was then calculated as:

$$Area\ (m^2)\ = \frac{(\pi\ *\ EDD^2\ *\ a)}{360} \qquad (3)$$

where the angle of view, *a*, for the Reconyx PC900 Professional Hyperfire cameras used in this study was 40° (Reconyx, 2017). We expected EDD and area of the field-of-view to vary by species, habitat type, and season because of the effects of snowpack, leaf phenology, and the thermal environment for infrared sensors (Hofmeester et al., 2019). To test for this variation, we considered eight broad habitat types (upland coniferous forest, upland deciduous forest, grassland, shrub, lowland forest, lowland grass, water, and human footprint) and two seasonal periods ('summer' between April 16 and October 15, and 'winter' being the rest of the year). We used a BIC-weighted model averaging framework to compare models of EDD by habitat type and season for each species or species group and obtain EDD estimates (Appendix S1).

For moose, EDD was generally estimated to be 6.5 m (90% CI: 6.32 to 6.67 m) across all habitat types and both seasons, producing an area of 14.7 $m^2$ (90% CI: 13.9 to 15.5 $m^2$) for the camera field-of-view. Other species had EDD that varied by habitat type, typically with longer EDD in more open habitats (Appendix S1). The EDD for moose may be underestimated because moose spend considerable time investigating the camera (discussed further under Assumption 2). Although investigating images were not used in the EDD calculation, investigative behavior may inflate the number of non-investigative images near the camera.

Using a single pole to create two distance bands was a minimal-effort approach to estimating EDD, but allowed us to collect enough data to compare estimated EDD across habitat types and seasons for multiple species. Directly measuring the distance and angle from the camera at first detection by tracking movement paths through the camera field-of-view provides a more refined EDD estimate, but requires significantly more effort (Rowcliffe et al., 2011). Using additional markers to delineate multiple distances in the field-of-view, as recommended by Hofmeester et al., (2017), would also contribute to a more finely delineated detection distance curve, but was not operationally feasible given our scale of deployment.

*Time in field-of-view ($T_F$).* Motion-activated cameras record animals as a series of discrete time-stamped images. To implement the TIFC approach, practitioners must convert these images into the total time the animal was in the field-of-view. Because we only collected still images, we needed to account for whether an animal left the field-of-view between two sequential images. Examining all images for evidence of the animal leaving or staying between images was too time intensive given the high volume of images collected. Instead, we examined sequential images from a subset of cameras to develop rules to apply to all images. For this sub-sample of images, we tagged whether the animal left the field-of-view in one image and returned in the next, or if it stayed in the field-of-view with no evidence of leaving in the interim. The classification was based on the position and behavior of the animal in the images before and after the interval., We found that for intervals of < 20 seconds between images, the animal had almost always stayed in the field-of-view, while the animal had almost always left the field-of-view for intervals > 120 seconds, regardless of species. For intervals between 20 and 120 seconds, we developed species-specific models of the probability that an animal left the field-of-view (Appendix S2).

For moose, we examined a random sample of 1,212 images. Of these, 30% of intervals of 20 seconds had evidence of the animal leaving the field-of-view and returning, rising to 80% for intervals of 120 seconds (Figure 1). Most other species showed similar relationships, with the notable exception that black bears (*Ursus americanus*) were only found to have left the field-of-view 40% of the time by the 120 second mark.

To convert discrete images into time in the camera field-of-view we:

1) Defined a "series" as consecutive images of a species with intervals < 120 seconds between any two consecutive images. An intermediate image without the species present ends the series, as do consecutive images showing an animal leaving and returning to the field-of-view (only applies to the sample of images tagged to support the development of the probability of leaving models). A series may range from a single image to hundreds of images.
2) Calculated total time in field-of-view for each series as the sum of time of all intervals < 20 seconds, plus the sum of intervals 20–120 seconds multiplied by (1 - probability of leaving) for that species and interval length. For example, if a series consists of three moose images separated by 5, 10, and 30 seconds, and the model for moose indicates there is a 40% chance that it would leave during a gap of 30 seconds, then the cumulative time for that series would be calculated as 5 + 10 + (30 × (1 – 0.4)) = 33 seconds.

3) We then account for time the animal is in the field-of-view before the first image and after the last image by adding to each series the time equivalent to the species-specific average number of seconds between consecutive images. For moose, this is 4.54 seconds. This additional time in the field-of-view is also added to series with a single image, which would otherwise have a time in the field-of-view of 0 seconds.
4) When multiple animals are simultaneously in the field-of-view, we use the average number of animals in images in the series as *N* in equation 1 for that series.

*Time camera is operating ($T_O$).* For most camera deployments, the total operating time is the time from initial set up to final collection. However, some cameras fail before recovery, most often because they run out of memory space or battery power, but sometimes because they are physically damaged. We programmed the cameras to also take time-lapse images every two hours to differentiate between cameras without animals in the field-of-view and cameras that had failed and identify timing of any failures. Cameras may also become displaced, and we consider the camera too displaced to use the images if the 5 m pole is no longer in the field-of-view or if the camera is tilted > 30° from horizontal, because these conditions greatly affect the EDD. We divided the time cameras are operating into two seasonal periods, summer (April 16 to October 15) and winter (October 16 to April 15), to account for changes in species seasonal movement patterns, habitat use, and detectability. The total numbers of days operating in each season were calculated.

*Calculating density.* Using equation 1, we calculated the density of each species at each camera. First, we calculated density separately for each of the two seasons, using the estimate of a species' time in the camera field-of-view during the season, the area of the camera field-of-view (based on seasonally-adjusted and habitat-specific EDD), and the camera operating time during that season. Next, we averaged the two seasonal estimates together for a yearly density estimate at each camera. For moose, the distribution of these estimates was extremely right-skewed, with the majority of cameras recording zero density (no detections), low densities at some cameras (one or a few individuals briefly passing by), and high densities at a small number of cameras (one or more individuals spending large amounts of time in front of the camera) (Figure 2). Of the 2,990 cameras used in this study, 838 recorded a moose detection.

*Testing assumptions of the TIFC method*

In this section, we examine three assumptions of TIFC required to estimate absolute density of a species. Failure to meet these assumptions will produce biased estimates of density. However, even if one or more of these assumptions is violated, the resulting estimates may still be useful as relative densities for comparisons in ecological applications, provided that the biases are consistent, or measurable and correctable, across the units of comparison. For instance, species' habitat models involve comparison between habitat types. Density estimates derived from camera data may still provide ecologically meaningful results for this purpose provided that these assumptions are either violated in equal ways across examined habitat types, or in measurable ways such that the different violations in different habitat types can be corrected.

*Assumption 1: Representative sampling of micro-habitats.* As with any quadrat sampling, the areas sampled must be a random, stratified random, or otherwise representative sample of the region for which inferences are being made (Fisher, 1925). At a landscape scale, our sampling design meets this expectation, as cameras are deployed in a random-systematic manner across the province of Alberta (ABMI, 2002). However, because individual cameras effectively sample a very small area (approximately 15-30 $m^2$), random sampling must also be considered at this "micro-habitat" scale. Because animals show strong preference for, or avoidance of, particular micro-habitats, non-random sampling could produce strongly biased estimates of absolute density that cannot be applied to the greater landscape.

In practice, cameras often cannot be placed in truly random micro-habitats. In forested areas, cameras are typically placed in areas that have an open field-of-view for 5m, because cameras facing tree trunks or thick patches of trees or shrubs are not useful for data collection. The selection for camera sites in open micro-habitats of forests will bias estimates of absolute density to the extent that animals prefer or avoid those more open areas within forests. Relative density for habitat modeling will be biased if animals show different degrees of preference or avoidance for more open areas in different forested habitat types (e.g., deciduous versus coniferous stands). We tested how TIFC density estimates of species differed between open versus densely treed micro-habitats in two upland forest types: deciduous (including deciduous-dominated mixedwood) and coniferous (ABMI, 2018). Using site photos, we classified the field-of-view of a random sample of camera deployments in each forest type as either typical treed forest (i.e. a field-of-view with vegetation density similar to that in the surrounding forest) or open (i.e. no trees within the 5m zone and the area being more open and vegetatively different from more distant surrounding forest). We further classified open sites as either productive (dense herb or shrub cover) or low-productive (sparsely vegetated), expecting that some species' preferences for openings might differ relative to productivity of the openings (Appendix S3). Of the 163 sampled deployments in deciduous stands, 105 were classified as open (64%), and of these open sites, 83 were productive (51% of total). A higher proportion of the 160 sampled deployments in coniferous stands were open, but of those 118 (74%) open sites, 51 (32% of total) were classified as low-productive. For each forest type, we calculated densities of common mammal species using the TIFC method described above in each of the three micro-habitat categories (typical, open-productive, and open-low productive), factoring out the effect of lure (See Incorporating Lure below). We estimated confidence intervals via bootstrapping with the camera deployment as the resampling unit.

For moose, cameras in productive openings of deciduous forest produced density estimates 1.62 times higher than those in typical treed micro-habitats (90% CI: 0.82-3.19), while low-productive openings provided density estimates 0.48 times as high as treed micro-habitats (90% CI: 0.13-1.18) (Figure 3, panel A). Productive openings in coniferous stands produced density estimates 3.60 times as high as typical treed sites (90% CI: 1.58-8.81), whereas low-productive openings in coniferous stands had densities similar to treed sites (Figure 3, panel B).

We cannot directly calculate the effect of these assumption violations, because we do not know what proportion of each broad stand type is composed of each micro-habitat. To illustrate the magnitude of these potential biases, we therefore use a hypothetical example in which low-productive and productive openings each occupy 5% of the total area of the two broad stand types. Using the results from Figure 3, we calculated the expected density estimates with representative sampling versus the proportions of opening-biased deployments in our sampling (64% in deciduous, 74% in coniferous). Our sampling biased towards open micro-habitats would result in density estimates for moose that are 1.24 times the density from representative sampling in deciduous forest, and 1.56 times greater in coniferous forest. Relative densities between the two broad forest types also change. With representative sampling, moose density in deciduous forest would be 2.49 times the density in coniferous forest, whereas biased sampling resulted in deciduous forest with estimated densities 1.97 times that in coniferous forest. This is based on the hypothetical example of 5% low-productive openings and 5% productive openings, and the true proportions of these openings may be higher or lower. To understand the exact magnitude of the bias introduced by selective sampling of open micro-habitats, further work is required to obtain sufficiently detailed landscape information. Advances in remote sensing may soon be able to provide the information required to properly calibrate these results, but current geospatial data are generally not available at the resolution required. Additional details of this test, including example images of the different treed/open categories, are presented in Appendix S3.

*Assumption 2: Movement not affected by the camera.* Most animal detections last only a few seconds as the animal crosses the camera field-of-view, with a small proportion lasting far longer. However, if animals spend even a few seconds investigating the camera or associated equipment (e.g. the 5m pole) on each visit, the total time in the field-of-view will be substantially inflated and result in an upwardly biased density estimate.

We assessed the overall proportion of time in the field-of-view that animals spent investigating the camera or 5m pole, including whether this proportion differed by broad habitat types (Figure 4). We measured investigative behaviors directly based on a subset of randomly selected series of images for each species. For moose, we selected 274 series, about 10% of the total., We classified each image in each series based on the behavior of the animal: (1) actively investigating or interacting with the pole or the camera, (2) behavior associated with investigation, including traveling directly towards the pole or the camera prior to investigating behavior, and/or lingering around the pole or camera after investigating, and (3) natural behaviors appearing to be unaffected by the pole or camera. Behavior 1 was generally unambiguous, but behavior 2 was more challenging to interpret, and it is unclear how much of the time animals spent in behavior 2 would have been spent in the field-of-view if they had not been attracted to the camera or pole.

Across all habitat types, moose spent 51% of their total time in the field-of-view investigating the camera or pole (behavior 1; 90% CI: 46-55%) (Figure 4, panel A). Proportion of time in behavior 1 was highest in grassland areas and lowest in deciduous forest. If investigating time was additive to time that moose would have otherwise been in the field-of-view, behavior 1 increased the overall density estimate by a factor of 2.02 (90% CI: 1.84-2.25), ranging from 1.51 to 3.41 across habitat types (with correspondingly wider confidence intervals). Combined, behaviors 1 and 2 represented 67% of total time in field-of-view (90% CI: 62-71%), with proportion of time highest in grass, shrub, and wet habitats, and lowest in deciduous forest, coniferous forest, and human footprint (Figure 4, panel B). Overall, including both behaviors 1 and 2 in the density calculations corresponded to a 3.00 times increase in density (90% CI: 2.62-3.47), with a range of 1.89 to 6.92 across habitat types (with correspondingly wider confidence intervals).

Among the species we detected, moose and black bears (55%) spent the highest proportion of time in behavior 1. Most other species spent 0–30% of their time in the field-of-view directly investigating (Appendix S4). For moose and black bears, violation of this assumption leads to a substantial overestimate of absolute density. The differences between habitat types also mean that the assumption was violated unequally, and that relative density estimates required by habitat modeling would be affected. Correction factors for investigation times can be used to correct for biases in time spent in front of the camera in different habitats. However, additional data collection would help address the wide uncertainty in these corrections, particularly if more habitat types needed to be included for detailed habitat modeling. It is also unclear if density corrections should be based strictly on direct investigation time, or whether the more ambiguous additional time for associated behaviors should also be included. Further details of this test, including results for other common species and example images for each behavior, are presented in Appendix S4.

*Assumption 3: Perfect detection near the camera.* Calculation of the camera EDD, which is used to define the area of a camera's field-of-view, assumes perfect detectability within 5 m of the camera. That is, when an animal is within 5 m of the camera, it is assumed that animal is always detected, and continues to be detected for the full duration of its stay in that area. To test this assumption, we conducted a paired camera experiment to measure the proportion of animal occurrences < 5 m from the camera that were accurately detected. We placed paired Reconyx PC900 Professional Hyperfire cameras at the reindeer (*Rangifer tarandus*) enclosure of the Edmonton Valley Zoo, which were the captive species that most closely matched moose in terms of body size and movement patterns. We set up one of the two cameras using the standard motion-triggered protocol described above, and we configured the second camera to take time-lapse images continuously at three second intervals. We mounted the cameras side-by-side, so that the time-lapse camera provided a (near) continuous record of the same area surveyed by the motion-triggered camera. Similar to the standard protocol, we placed a pole 5 m in front of the cameras. We used the images collected from both cameras to measure the proportion of reindeer occurrences within 5 m that were detected by the motion-triggered camera.

Compared to the number of occurrences captured by the time-lapse cameras, which we assume to be the true number, the motion-triggered camera detected 95% of these events. The missed detections all involved small parts of the animal body (e.g. antler tines) at the extreme periphery of the field-of-view, which may be a zone that is unreliably detected by the PC900 Professional Hyperfire camera model used (Apps and McNutt, 2018). These results suggest that species similar to reindeer in both body size and movement patterns (e.g. moose) are likely to be detected reliably within the 5m zone. Although not tested in our experiment, vegetation cover, snow depth, air temperature, movement speed, and body size may all impact the detection rate. Unreliable detection due to these factors has been identified as a concern with the PC900 Hyperfire cameras (Urbanek et al, 2019), so additional field experiments will be required to confirm whether this high rate of detection within 5 m in a zoo setting is comparable to field sampling or whether it holds for additional species.

*Incorporating Lure*

For many species, and carnivores in particular, detections at randomly located camera sites can be very low. Lure or other attractants are often used to increase the number of detections at cameras, which can help reduce the high inherent measurement error of cameras (Holinda et al., 2020). However, lures clearly violate the assumption that animal movement is not influenced by the camera deployment. Furthermore, differential attraction to lures in different habitats would bias habitat models based on camera data.

We deployed both lured and unlured cameras in a paired design to calibrate lure effects for each species. As described in the Data Collection section above, each site used four cameras spaced 600-m apart, two of which were lured. This design allowed for a simple paired comparison of species' occurrences and time in field-of-view. We examined results for common species at 558 core sites from both lured and unlured cameras (992 of each, with two pairs at most sites) with similar total operating times. We summarized the ratio of lured:unlured results by species in three measures: a) occurrence (presence/absence at the camera over the entire deployment time); (b) density given occurrence (density at only cameras where the species was present); and (c) total density (the product of occurrence and density given occurrence, i.e. our density estimates as described above). We used bootstrapping to calculate confidence intervals for each mean ratio, with site as the resampling unit.

The ratio of the mean lured:unlured values for moose occurrence was 1.07 (90% CI: 0.99-1.16), 1.17 (0.92-1.5) for density given occurrence, and 1.26 (0.98-1.62) for total density. We found more substantial positive effects of lure for other species, particularly fisher (*Pekania pennanti*), red fox (*Vulpes vulpes*), and grey wolf (*Canis lupus*) (Appendix S5). For these carnivore species, a larger proportion of the effect on total density came from the density given occurrence component, indicating that lure was primarily effective for increasing the time animals spend in the camera field-of-view. Occurrence increases due to lure tended to be smaller, which suggests that animals are not being drawn in from large distances. We use the total density ratios to correct estimates of densities at lured cameras to an unlured density equivalent.

**Application – Moose Density in Wildlife Management Units**

We used moose images collected from the remote cameras to illustrate regional density estimation using the TIFC method and to compare TIFC and aerial survey density estimates. Aerial surveys are currently the primary approach used by wildlife managers for monitoring moose populations in the province of Alberta. Wildlife Management Units (WMUs) are key spatial units for wildlife management decisions in the province, such as establishing hunting quotas and determining priority areas for recovery actions. Since 2014, the Alberta provincial government has used distance sampling (Buckland et al., 2001; Thomas et al., 2002) on aerial ungulate surveys to estimate moose densities in most WMUs (Peters et al., 2014). Distance transect surveys are flown in winter, with observers recording the perpendicular distance from the transect to the observed moose. Moose density in the WMU and associated precision of this estimate is calculated with Distance software (Thomas et al., 2010). The distance method assumes that animals are detected with certainty along the transect line and that distances are measured without error (Buckland et al., 2001). To the extent that these assumptions are met, the results of aerial surveys can be considered an unbiased estimate of moose density in each WMU. We obtained moose density estimates, including confidence intervals, from reports available on the Alberta Environment and Parks website (AEP 2021). We used only estimates based on distance sampling and restricted our sample to aerial surveys conducted between 2014 and 2020 in the boreal region of the province (Figure 5). For two WMUs with more than one survey completed during this time period, we used the aerial survey density estimate that was closest in time to the camera sampling done in those WMUs.

To calculate mean moose density for each WMU using the TIFC method, we used camera data from the core sites described previously, as well as additional data from five 4 × 12-km grids of 25 randomly placed cameras each (minimum spacing 1 km) (Figure 5). The same camera models and set-up protocol were used at all deployments, including a reference pole at 5 m. We only calculated density estimates for WMUs with at least 15 cameras deployed between 2014 and 2020. Confidence intervals were estimated as a zero-inflated log-normal distribution using Monte Carlo simulation of both the presence/absence (binomial) and density given occurrence (log-normal) components. We did not attempt to match years of aerial surveys with years of camera sampling with WMUs because this would severely limit sample size; however, the majority of camera data were collected within two years of the corresponding aerial survey. A total of 29 WMUs were used in this comparison, ranging in size from 1,917 to 21,463 $km^2$ (Figure 5). Moose density estimates from both the aerial surveys and cameras for each WMU are listed in Appendix S6.

To compare the two methods, we fit a linear regression of camera density as a function of the aerial survey density (without intercept) in R v.4.0.4 (R Core Team 2021). Because of the number of cameras per WMU was highly variable (ranging from 15 to 217), we weighted the WMUs in inverse proportion to the square root of the number of cameras (as a proxy for the precision of camera estimate). Camera estimates were positively related to aerial survey estimates across WMUs ($r^2 = 0.84$), but with wide uncertainty at the level of an individual WMU (Figure 6). On average, camera-derived moose density estimates were 2.47 times higher than aerial survey estimates (90% CI: 2.13 - 2.81).

We expected higher estimates of moose density based on camera data because of the assumption violations we documented: moose were more abundant in microhabitat forest openings that our cameras sampled selectively and they spent high proportions of time investigating the camera and pole. We cannot correct for the former violation, because we do not know how much our cameras overrepresented openings. However, we can correct for investigating time. Using the measured direct investigation times by habitat type (Figure 4) and the known habitat types of each camera used in the WMU estimates, we removed the estimated time moose spent investigating the camera and pole, and re-calculated densities for each camera. With this adjustment, density estimates from camera traps were 1.3 times as high as estimates from aerial surveys (90% CI: 1.1-1.5) (Figure 7). Higher initial density estimates from cameras in WMUs may have largely been due to this bias from moose investigating cameras. However, there was still a significant amount of uncertainty in the corrected relationship (which does not include additional uncertainty from the correction factor itself) and wide variation among individual WMUs.

*Data Accessibility*

All data and code used for these analyses, including the assumptions testing, are available at *https://github.com/mabecker89/tifc-method*. The zipped folder, DataS1.zip, contains these code files as R scripts.

## Discussion

In this paper we describe the implementation of the TIFC model to estimate animal density from motion-activated cameras. We show how the parameters needed to implement the model – effective detection distance and time in field-of-view – can be estimated as part of a camera-based monitoring program, and how important assumptions can be tested using typical field-collected data or with little additional effort.

Quadrat sampling, which underlies the TIFC method, is a familiar data collection approach for most ecologists. However, unlike most quadrat surveys, camera quadrats are tiny compared to the ranges of animals they are used to monitor (a few square meters for a camera field-of-view versus many hectares or square kilometers for some mammal ranges). In our WMU camera data, there was an average of $7.3 \times 10^{-6}$ moose in any one camera field-of-view at any given time. The actual density of a species in the field-of-view is almost always 0, but it is extremely high for the few seconds that an animal is present. Camera field-of-views are able to work as quadrats because TIFC integrates extremely local density estimates over long periods of time. However, this aspect of camera sampling makes density estimates from cameras particularly sensitive to violations of assumptions and subject to high measurement error, much more so than in more familiar quadrat applications.

Because of the small scale of the camera field-of-view, non-representative sampling is a risk, even if only due to practical constraints on camera placement. Cameras placed in small forest openings can estimate moose densities up to three times as high as treed patches, reflecting small-scale movement or foraging decisions by moose. In this case, we cannot meet the assumption of truly random sampling, but we can correct for the bias, if we can collect auxiliary information on the prevalence of small openings in different stand types. Emerging remote-sensing technology may help address this limitation in the future. The potential biases would be more pronounced if cameras were intentionally placed in micro-sites with high animal abundance, such as along game trails, ridges, or edges of water bodies. The small field-of-view also means that animals typically only spend a few seconds in front of the camera, such that additional time spent investigating the camera equipment can substantially increase density estimates, as we demonstrated with moose.

Completely avoiding assumption violations is not possible practically, and it is therefore important to measure and correct for them. We have shown that the detailed information provided by images can be used to test many of the critical assumptions inherent in camera sampling. Examining the images can also identify other potential biases and ways to account for them. For example, images provide evidence that certain species follow the tracks of field crews into the camera field-of-view shortly after winter camera deployment, effectively attracting animals to the camera and inflating densities. We reduce the bias by having long deployment times, but we could also use these images to directly measure that behavior and how much it affects density compared to times later in the deployment when snow has filled in the tracks left by field crews. Preservation of much useful information for testing assumptions is a benefit of camera surveys.

Good design and direct testing can reduce and correct biases from many violations of assumptions, but we doubt that they can ever be eliminated in camera sampling. Absolute density estimates will always be affected to some extent. Even relative densities, sufficient for many ecological applications, require tests and corrections for differences in how assumptions are violated among cameras or groups of cameras being compared. We demonstrated this for broad habitat types, but additional work is needed to apply corrections (e.g., measuring prevalence of small openings), as well as to collect additional information for more detailed habitat types when finer habitat models are an objective. Additionally, monitoring programs are often directed towards long-term monitoring of trends in species' populations. Using camera-based data for that purpose requires testing that violations of assumptions do not change over time, including as the landbase changes through development or natural processes. For example, increased fires or human-created openings may provide more browsing opportunities for moose, possibly reducing their preference for small forest openings (i.e. a function response). We cannot anticipate the effects of all such changes, and therefore recommend that ongoing assumption testing be built into long-term monitoring programs.

A high level of measurement error is an additional challenge for camera-based monitoring with the TIFC method. In our moose example, WMUs had an average of 52 cameras and 107 total moose-minutes in those cameras' fields-of-view. A single animal resting in front of a single camera for 90 minutes would almost double the density estimate for a typical WMU. The highly skewed distribution of density estimates from individual cameras (Figure 2) is a manifestation of the effect of the small quadrat size of camera fields-of-view. Large numbers of cameras, or longer deployment times, are therefore needed to attain precise estimates (Figure 8). Alternative sampling strategies that violate assumptions may be viable options if calibrations can be developed, as we showed with our operational calibration of paired lured and unlured cameras. Lure increases detections of carnivores, which should increase precision, although this benefit needs to be balanced against the additional uncertainty introduced by the lure calibration estimate. For long-term monitoring, we also need to be aware that the effects of lures can change over time, which means we need to continue operational calibration to measure and compensate for any such changes. Placing some cameras in high-use areas, such as game trails, may also be an option for reducing uncertainty, as long as cameras are deployed in a paired design that allows calibration to sites more representative of the larger area. Because trail use may also change over time – for example, as development creates more alternative travel routes in the region – that calibration would also need to be replicated in different landscape contexts and over time.

Despite the challenges of sensitivity to violations of assumptions and high measurement error, cameras can provide useful alternative or complementary estimates of mammal abundance. In our example of moose densities in WMUs of Alberta, that the relationship was close to 1:1 with aerial surveys after correcting for one assumption violation was probably fortuitous, given the wide uncertainty in individual WMU estimates, lack of correction for at least one other known bias, as well as uncertainty and potential biases in the aerial surveys themselves (Caughley, 1974; Anderson and Lindsey, 1996; Oyster et al, 2018). Additionally, the camera sites were not chosen to systematically sample a WMU (e.g. upland areas versus lowland areas, or in relation to human disturbance), thus the representativeness of camera sampling for each WMU may not be complete, particularly for WMUs with fewer cameras. Nonetheless, the cameras did capture the same variation across the range of WMUs as the aerial surveys, with similar absolute densities overall. We are not recommending that cameras replace aerial ungulate surveys, but a full comparison of the two methods, including costs, risks, and benefits across a range of situations would help determine if there are circumstances where there are advantages to one or the other. One important consideration is that cameras can provide density estimates for a variety of mammal species, beyond those captured by aerial surveys. Well-designed and rigorously-tested camera surveys are likely the only feasible option for concurrently monitoring a range of medium-sized to large mammal species.

## Acknowledgements

The Alberta Biodiversity Monitoring Institute provided all field-related support. The collection and analysis of the data in this project was funded by the Regional Industry Caribou Collaboration (RICC), Alberta Environment and Parks (AEP), and the Oil Sands Monitoring Program (OSM). It is independent of any position of the OSM Program. The authors thank Simon Slater and Andrew Braid from AEP for their help retrieving and interpreting the aerial survey data, as well as Paul Williams for his assistance and advice in setting up the paired camera experiment at the Edmonton Valley Zoo.

## Literature Cited

ABMI [Alberta Biodiversity Monitoring Institute]. 2018. Alberta wall-to-wall vegetation layer including "backfilled" vegetation in human footprints (Version 6). Alberta, Canada. http://www.abmi.ca

AEP [Alberta Environment and Parks]. 2021. Aerial wildlife survey reports. Alberta, Canada. https://www.alberta.ca/aerial-wildlife-survey-reports.aspx

Amin, R., A. E. Bowkett, and T. Wacher. 2016. The use of camera-traps to monitor forest antelope species. Pages 190–216 *in* Antelope Conservation: From Diagnosis to Action. John Wiley & Sons, Ltd, Chichester, UK.

Anderson, C. R., and F. G. Lindzey. 1996. Moose sightability model developed from helicopter surveys. Wildlife Society Bulletin 24:247–259.

Apps, P., and J. W. McNutt. 2018. Are camera traps fit for purpose? A rigorous, reproducible and realistic test of camera trap performance. African Journal of Ecology 56:710–720.

Bessone, M., H. S. Kühl, G. Hohmann, I. Herbinger, K. P. N'Goran, P. Asanzi, P. B. Da Costa, V. Dérozier, E. D. B. Fotsing, B. I. Beka, M. D. Iyomi, I. B. Iyatshi, P. Kafando, M. A. Kambere, D. B. Moundzoho, M. L. K. Wanzalire, and B. Fruth. 2020. Drawn out of the shadows: Surveying secretive forest species with camera trap distance sampling. Journal of Applied Ecology 57:963–974.

Buckland, S. T. 1987. On the variable circular plot method of estimating animal density. Biometrics 43:363-384.

Buckland, S. T., E. A. Rexstad, T. A. Marques, and C. S. Oedekoven. 2015. Distance Sampling: Methods and Applications. Springer International Publishing, New York, New York, USA.

Buckland, S. T., D. R. Anderson, K. P. Burnham, J. L. Laake, D. L. Borchers, and L. Thomas. 2001. Introduction to Distance Sampling: Estimating Abundance of Biological Populations. Oxford University Press, Oxford, UK.

Burton, A. C., E. Neilson, D. Moreira, A. Ladle, R. Steenweg, J. T. Fisher, E. Bayne, and S. Boutin. 2015. Wildlife camera trapping: A review and recommendations for linking surveys to ecological processes. Journal of Applied Ecology 52:675–685.

Caravaggi, A., M. Zaccaroni, F. Riga, S. C. Schai-Braun, J. T. A. Dick, W. I. Montgomery, and N. Reid. 2016. An invasive-native mammalian species replacement process captured by camera trap survey random encounter models. Remote Sensing in Ecology and Conservation 2:45–58.

Caughley, G. 1974. Bias in Aerial Survey. The Journal of Wildlife Management 38:921.

Chandler, R. B., and J. Andrew Royle. 2013. Spatially explicit models for inference about density in unmarked or partially marked populations. Annals of Applied Statistics 7:936–954.

Cusack, J. J., A. Swanson, T. Coulson, C. Packer, C. Carbone, A. J. Dickman, M. Kosmala, C. Lintott, and J. M. Rowcliffe. 2015. Applying a random encounter model to estimate lion density from camera traps in Serengeti National Park, Tanzania. Journal of Wildlife Management 79:1014–1021.

Fisher, R.A. 1925. Statistical Methods for Research Workers. Oliver and Boyd, London, UK.

Garland, L., Neilson, E., Avgar, T., Bayne, E., & Boutin, S. 2020. Random encounter and staying time model testing with human volunteers. The Journal of Wildlife Management, 84: 1179-1184.

Gilbert, N. A., J. D. J. Clare, J. L. Stenglein, and B. Zuckerberg. 2021. Abundance estimation of unmarked animals based on camera-trap data. Conservation Biology 35:88–100.

Hofmeester, T. R., J. P. G. M. Cromsigt, J. Odden, H. Andren, J. Kindberg, and J. D. C. Linnell. 2019. Framing pictures: A conceptual framework to identify and correct for biases in detection probability of camera traps enabling multi-species comparison. Ecology and Evolution 9:2320–2336.

Hofmeester, T. R., J. M. Rowcliffe, and P. A. Jansen. 2017. A simple method for estimating the effective detection distance of camera traps. Remote Sensing in Ecology and Conservation 3:81–89.

Holinda, D., J. M. Burgar, and A. C. Burton. 2020. Effects of scent lure on camera trap detections vary across mammalian predator and prey species. PLoS ONE 15:e0229055.

Howe, E. J., S. T. Buckland, M. Després-Einspenner, and H. S. Kühl. 2017. Distance sampling with camera traps. Methods in Ecology and Evolution 8:1558–1565.

Karanth, K. U., and J. D. Nichols. 1998. Estimation of tiger densities in India using photographic captures and recaptures. Ecology 79:2852–2862.

Laurent, M., M. Dickie, M. Becker, R. Serrouya, and S. Boutin. 2021. Evaluating the Mechanisms of Landscape Change on White-Tailed Deer Populations. The Journal of Wildlife Management 85:340–353.

Mace, G. M., N. J. Collar, K. J. Gaston, C. Hilton-Taylor, H. R. Akçakaya, N. Leader-Williams, E. J. Milner-Gulland, and S. N. Stuart. 2008. Quantification of extinction risk: IUCN's system for classifying threatened species. Conservation Biology 22:1424–42.

Manly, B. 2014. Standard Sampling Methods and Analyses. Pages 22-47 *in* Introduction to Ecological Sampling. Chapman and Hall / CRC, New York, New York, USA.

Mills, L. S. 2007. Conservation of Wildlife Populations: demography, genetics, and management. Blackwell Publishing Ltd, Malden, Massachusetts, USA.

Moeller, A. K., P. M. Lukacs, and J. S. Horne. 2018. Three novel methods to estimate abundance of unmarked animals using remote cameras. Ecosphere 9:e02331.

Nakashima, Y., K. Fukasawa, and H. Samejima. 2018. Estimating animal density without individual recognition using information derivable exclusively from camera traps. Journal of Applied Ecology 55:735–744.

Oyster, J. H., I. N. Keren, S. J. K. Hansen, and R. B. Harris. 2018. Hierarchical mark-recapture distance sampling to estimate moose abundance. Journal of Wildlife Management 82:1668–1679.

Palmer, M. S., A. Swanson, M. Kosmala, T. Arnold, and C. Packer. 2018. Evaluating relative abundance indices for terrestrial herbivores from large-scale camera trap surveys. African Journal of Ecology 56:791–803.

Peters, W., M. Hebblewhite, K. G. Smith, S. M. Webb, N. Webb, M. Russell, C. Stambaugh, and R. B. Anderson. 2014. Contrasting aerial moose population estimation methods and evaluating sightability in west-central Alberta, Canada. Wildlife Society Bulletin 38:639–649.

R Core Team (2021). R: A language and environment for statistical computing. R Foundation for Statistical Computing, Vienna, Austria. URL https://www.R-project.org/.

Rayan, D. M., S. W. Mohamad, L. Dorward, S. A. Aziz, G. R. Clements, W. C. T. Christopher, C. Traeholt, and D. Magintan. 2012. Estimating the population density of the Asian tapir (*Tapirus indicus*) in a selectively logged forest in Peninsular Malaysia. Integrative Zoology 7:373–380.

Rovero, F., and A. R. Marshall. 2009. Camera trapping photographic rate as an index of density in forest ungulates. Journal of Applied Ecology 46:1011–1017.

Rowcliffe, J. M., P. A. Jansen, R. Kays, B. Kranstauber, and C. Carbone. 2016. Wildlife speed cameras: measuring animal travel speed and day range using camera traps. Remote Sensing in Ecology and Conservation 2:84–94.

Rowcliffe, J. M., C. Carbone, P. A. Jansen, R. Kays, and B. Kranstauber. 2011. Quantifying the sensitivity of camera traps: An adapted distance sampling approach. Methods in Ecology and Evolution 2:464–476.

Rowcliffe, J. M., J. Field, S. T. Turvey, and C. Carbone. 2008. Estimating animal density using camera traps without the need for individual recognition. Journal of Applied Ecology 45:1228–1236.

Royle, J. A., J. D. Nichols, K. U. Karanth, and A. M. Gopalaswamy. 2009. A hierarchical model for estimating density in camera-trap studies. Journal of Applied Ecology 46:118–127.

Steenweg, R., M. Hebblewhite, R. Kays, J. Ahumada, J. T. Fisher, C. Burton, S. E. Townsend, C. Carbone, J. M. Rowcliffe, J. Whittington, J. Brodie, J. A. Royle, A.

Switalski, A. P. Clevenger, N. Heim, and L. N. Rich. 2017. Scaling-up camera traps: monitoring the planet's biodiversity with networks of remote sensors. Frontiers in Ecology and the Environment 15:26–34.

Thomas, L., S. T. Buckland, E. A. Rexstad, J. L. Laake, S. Strindberg, S. L. Hedley, J. R. B. Bishop, T. A. Marques, and K. P. Burnham. 2010. Distance software: Design and analysis of distance sampling surveys for estimating population size. Journal of Applied Ecology 47:5–14.

Thomas, L., S. T. Buckland, K. P. Burnham, D. R. Anderson, J. L. Laake, D. L. Borchers, and S. Strindberg. 2002. Distance Sampling. Pages 544–552 *in* A. H. El-Shaarawi and W. W. Piegorsch, editors. Encyclopedia of Environmetrics Volume 1, A-D. John Wiley and Sons, Ltd., Chichester, UK.

Trolle, M., A. Noss, J. Cordeiro, and L. Oliveira. 2008. Brazilian tapir density in the Pantanal: A comparison of systematic camera-trapping and line-transect surveys. Biotropica 40:211–217.

Urbanek, R. E., Ferreira, H. J., Olfenbuttel, C., Dukes, C. G., & Albers, G. 2019. See what you've been missing: An assessment of Reconyx® PC900 Hyperfire cameras. Wildlife Society Bulletin, 43: 630-630.

Warbington, C. H., and M. S. Boyce. 2020. Population density of sitatunga in riverine wetland habitats. Global Ecology and Conservation 24:e01212.

Wilson, G. J., and R. J. Delahay. 2001. A review of methods to estimate the abundance of terrestrial carnivores using field signs and observation. Wildlife Research 28:151–164.

Zero, V. H., S. R. Sundaresan, T. G. O'Brien, and M. F. Kinnaird. 2013. Monitoring an Endangered savannah ungulate, Grevy's zebra *Equus grevyi*: Choosing a method for estimating population densities. Oryx 47:410–41.

**Figures**

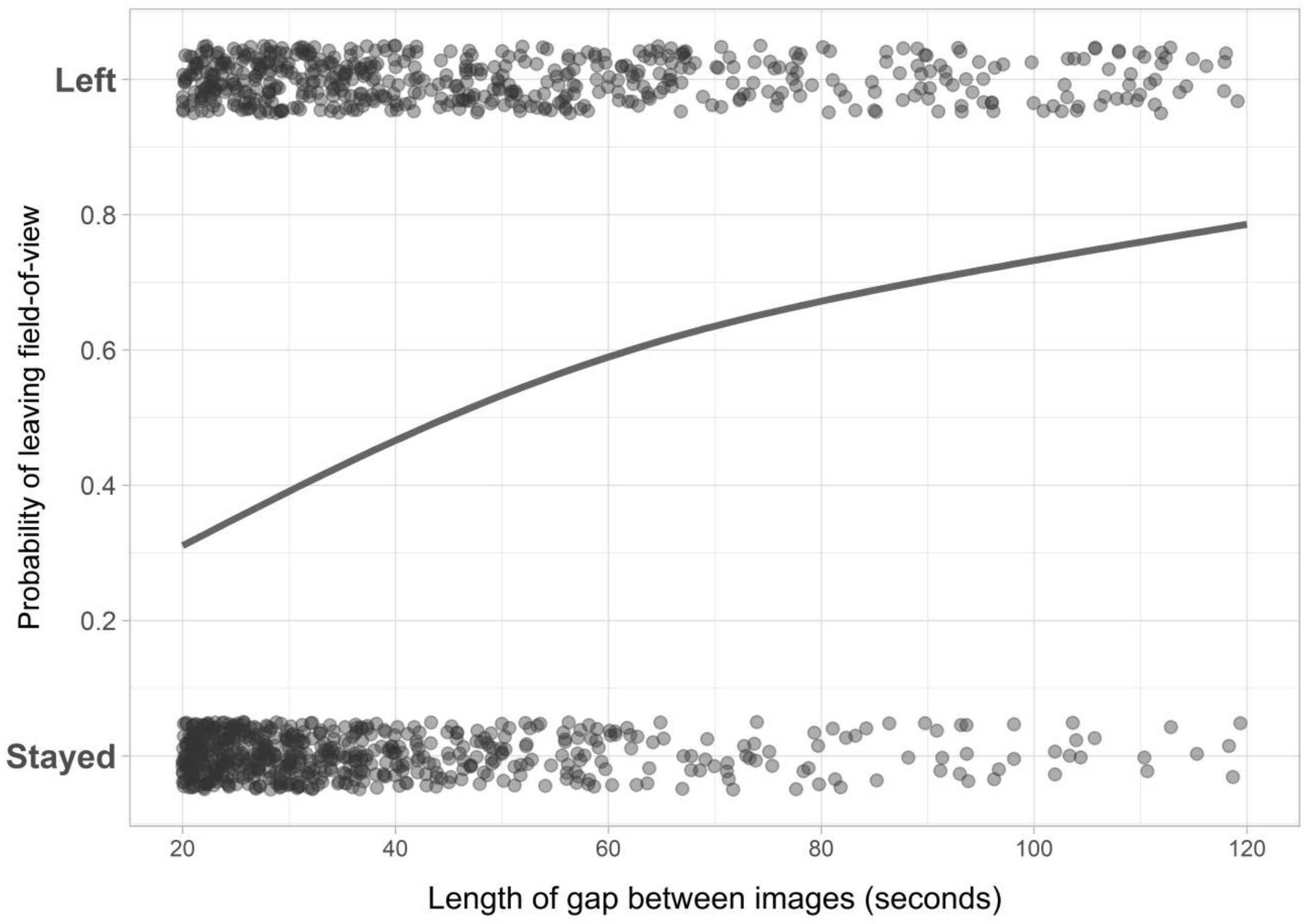

*Figure 1*. Estimated probability of moose leaving the camera field-of-view based on the length of the interval between sequential images. A total of 1,212 images were sampled to construct this model.

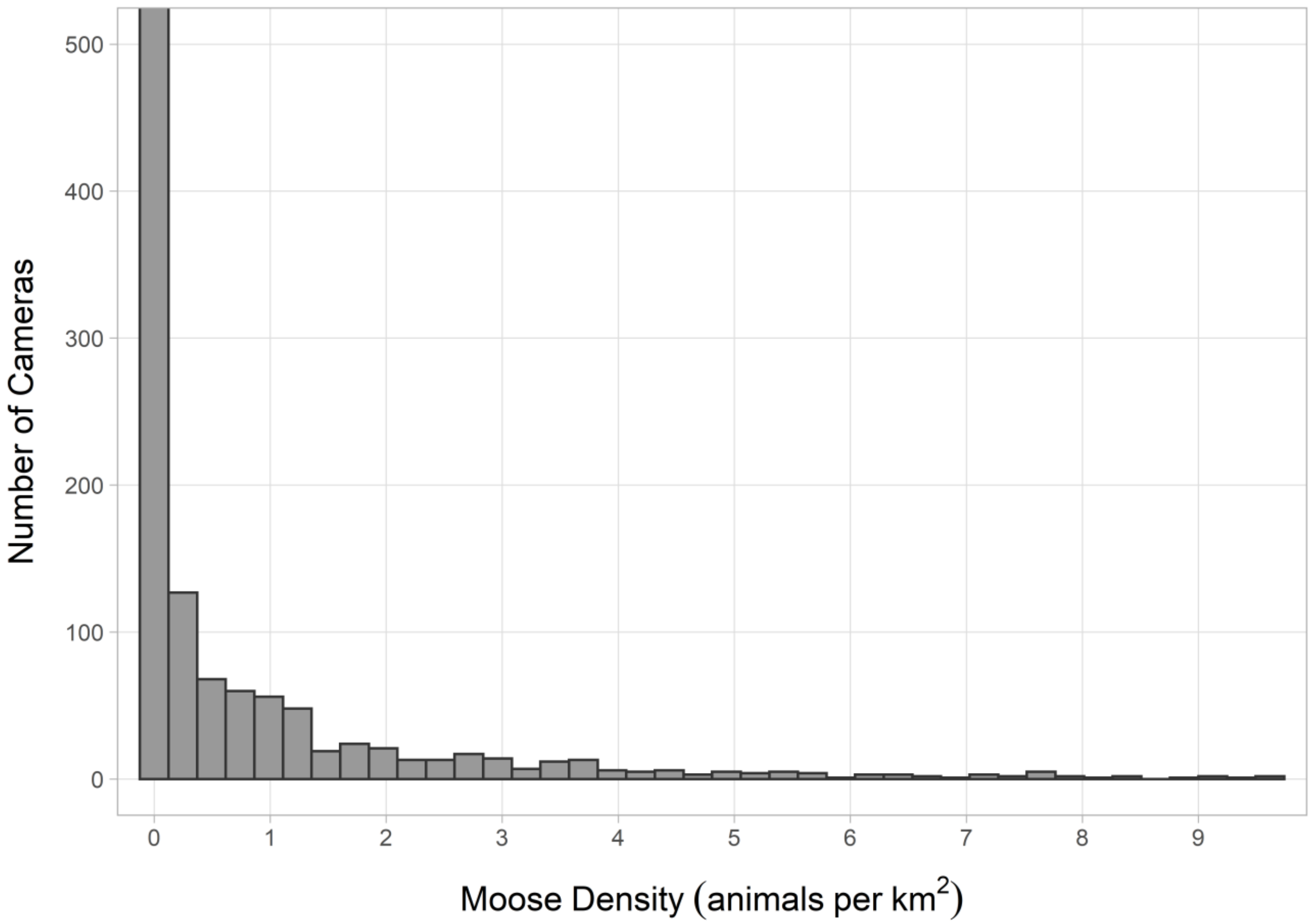

*Figure 2.* Distribution of calculated moose densities across the 2,990 cameras. 17 cameras with density values above 10 animals per $km^2$ were omitted from the figure (max of 52 animals per $km^2$). The leftmost bar, containing density values between 0 and 0.12, consisted of 2,232 cameras, but the figure was truncated at *c.* 500 to more clearly visualize bars with lower numbers of cameras. Note that these densities are accurate only to the extent that the assumptions discussed below are met.

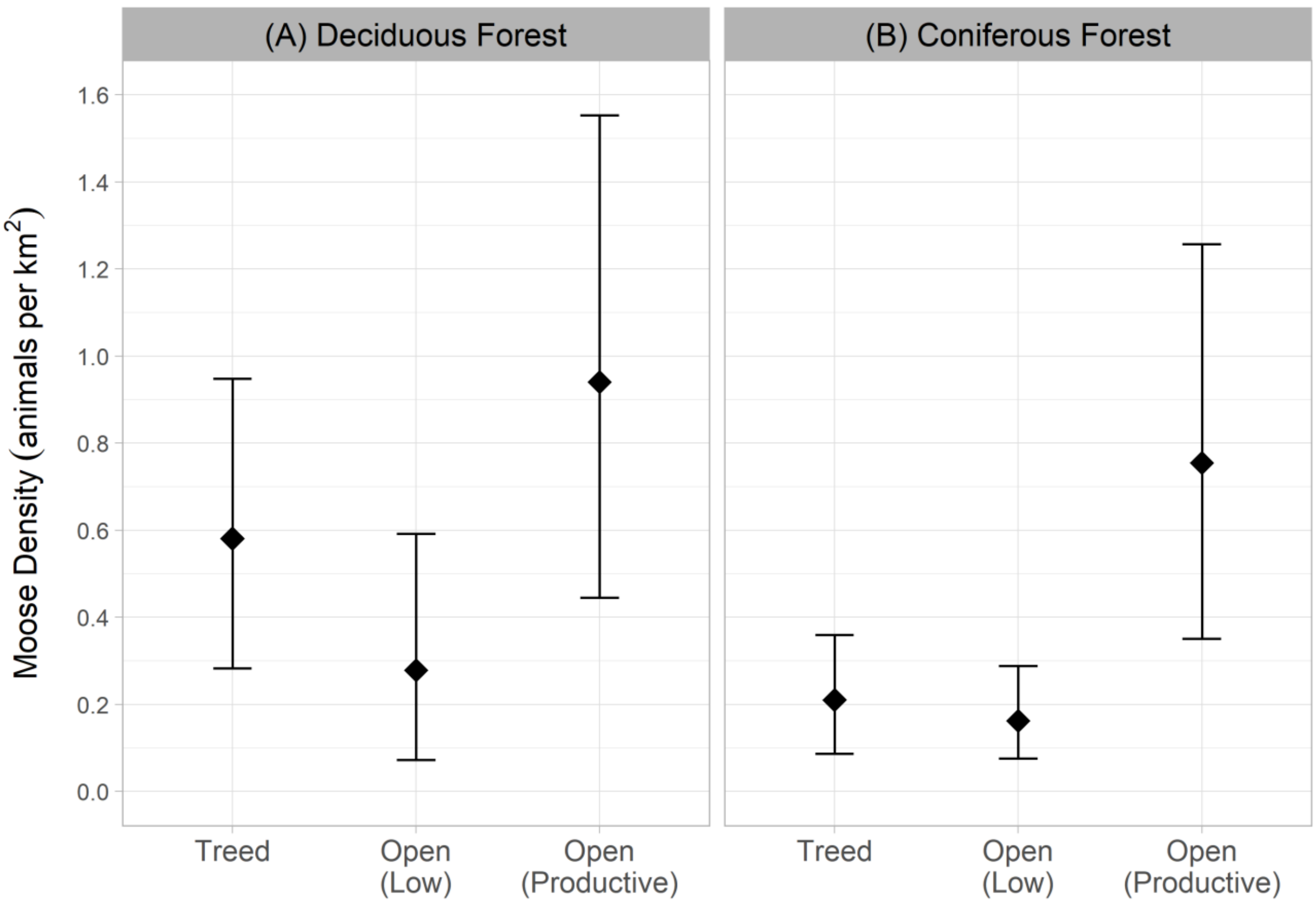

*Figure 3.* Average moose densities at deployments located in typical treed vs open (low-productive (low) and productive) microhabitats in both deciduous (panel A) and coniferous (panel B) forest stands. Error bars reflect 90% confidence intervals.

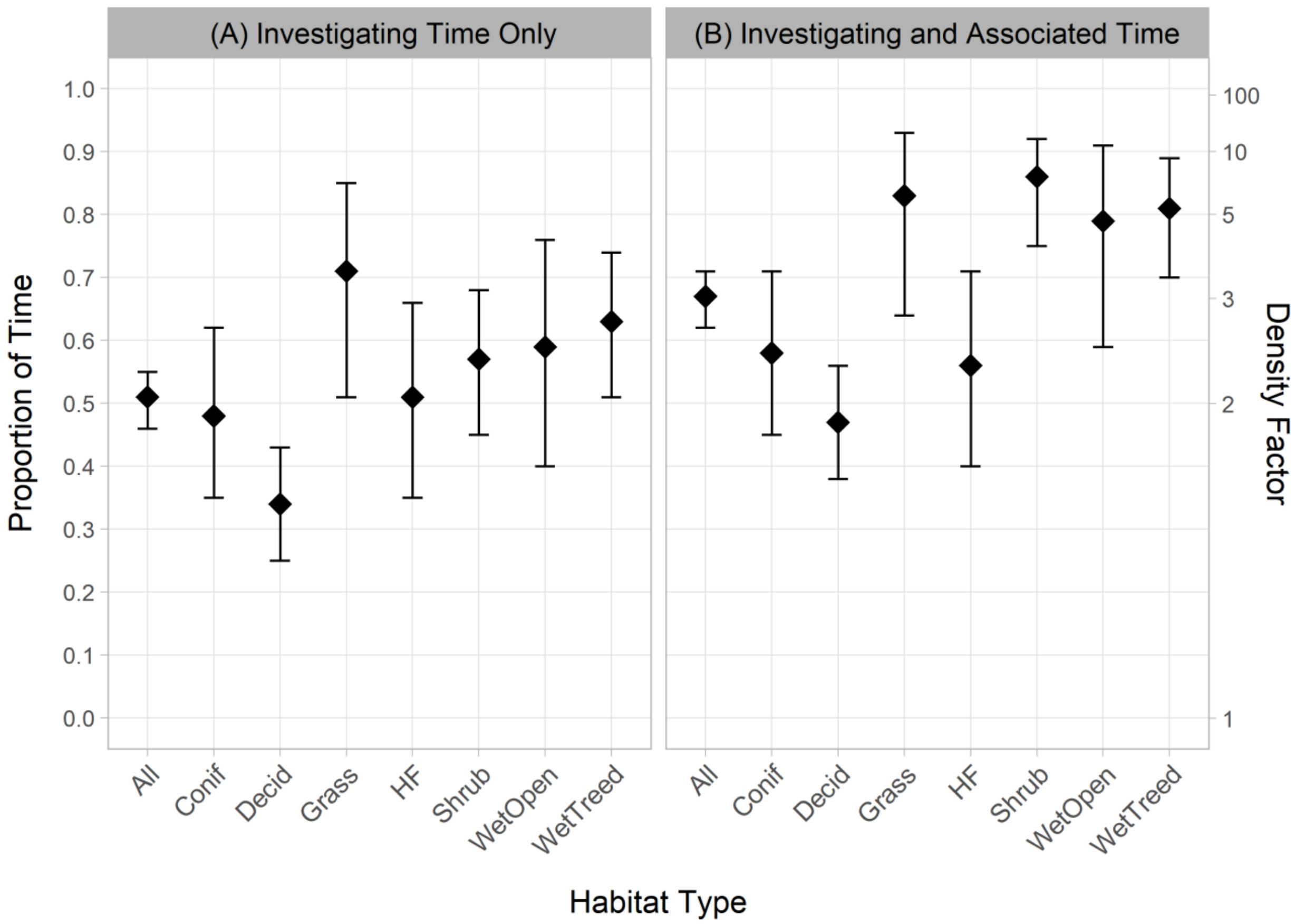

*Figure 4.* Proportion of time in front of the camera that moose spend investigating the camera and pole (panel A; behavior 1), and time spent investigating plus associated behaviors (panel B; behaviors 1 and 2). Density factor is the corresponding increase to downstream density estimates, based on the additional time spent in the behaviors. Error bars represent 90% confidence intervals.

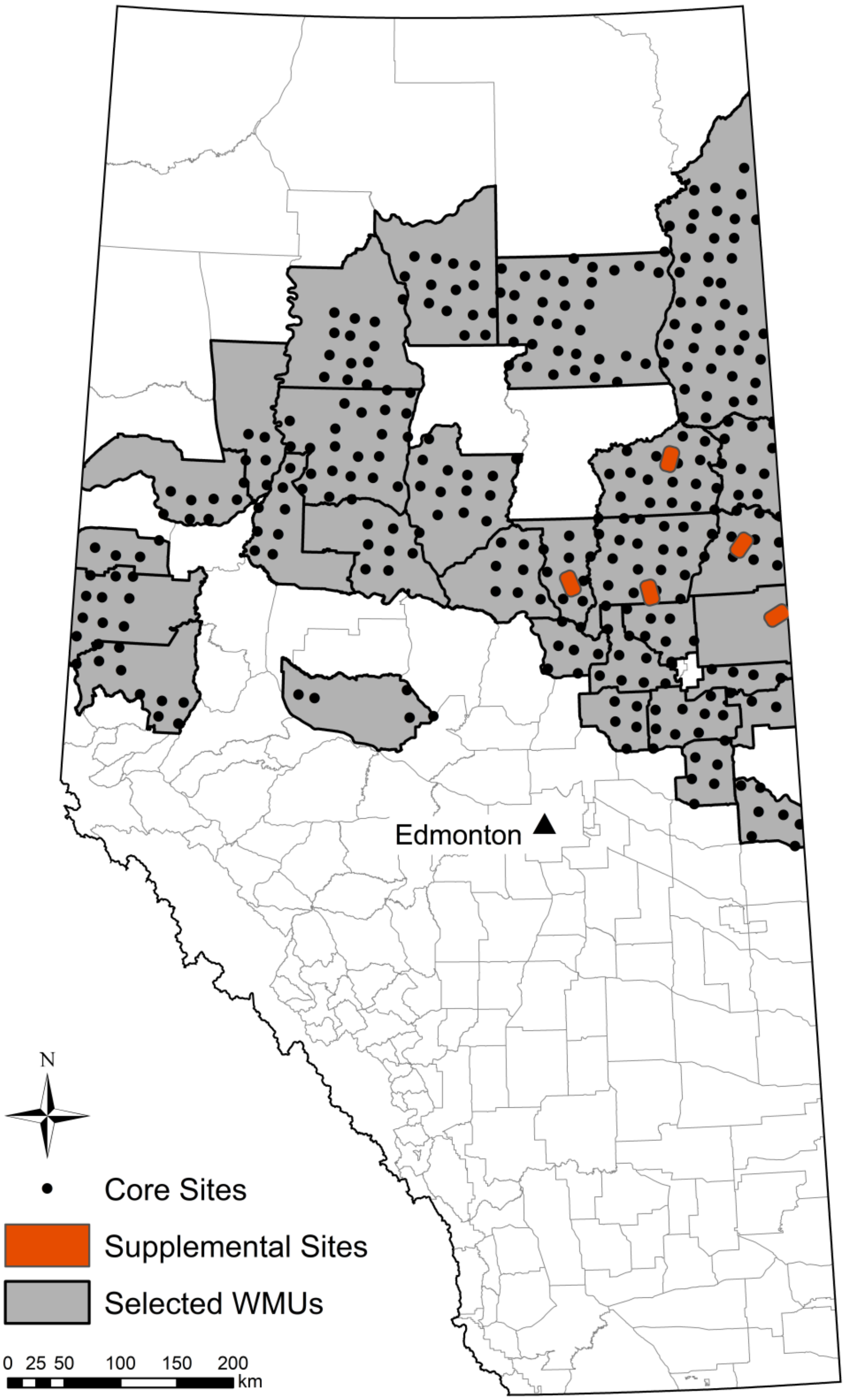

*Figure 5*. Location of Wildlife Management Units (WMUs) with available data from at least 15 camera deployments (2015-2020) and aerial surveys (2014-2020) within Alberta, Canada. Core sampling sites have 4 cameras placed in a square, separated by 600 m. Supplemental sites have 25 cameras with a minimum 1-km spacing within a 4 × 12 km area.

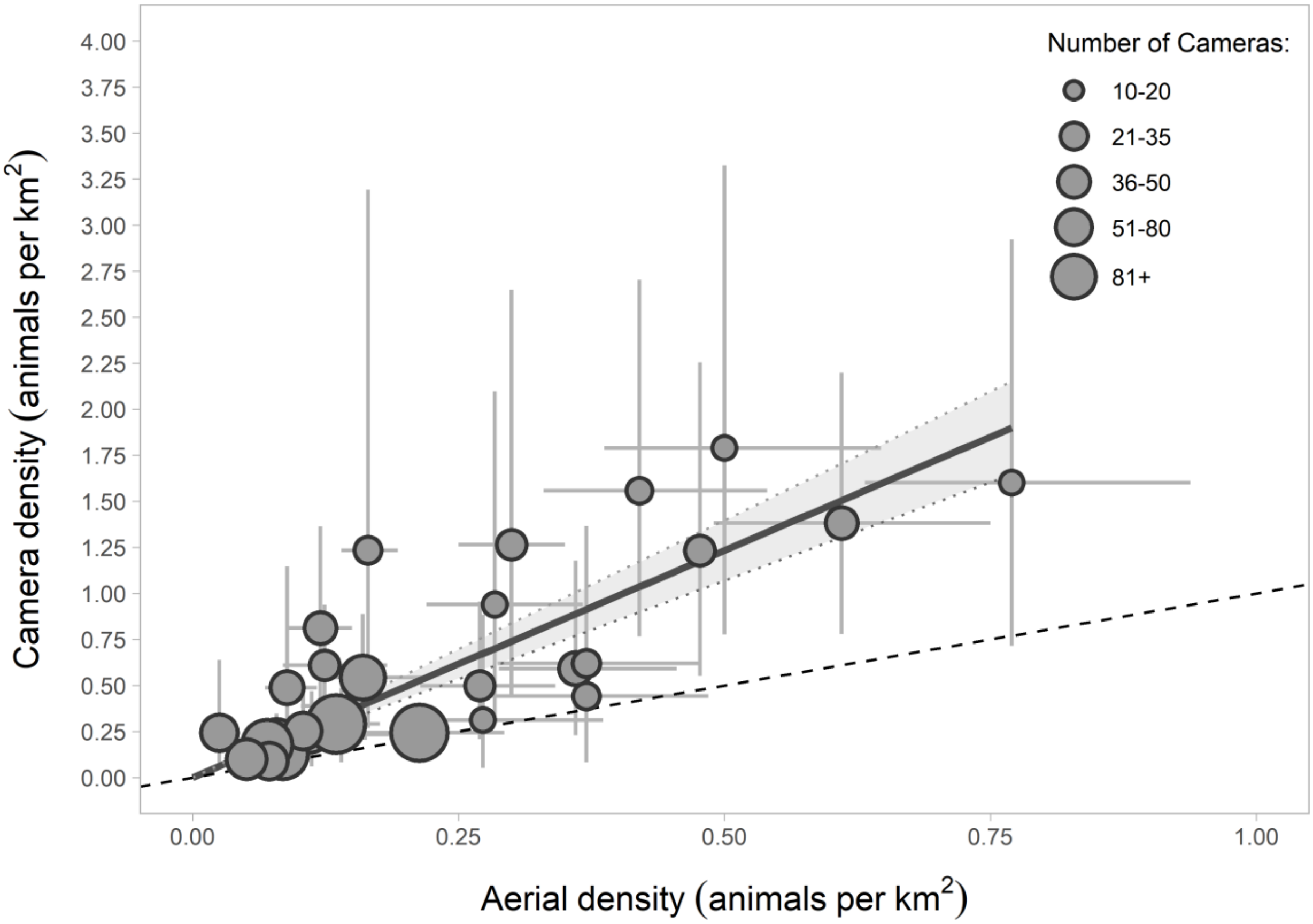

*Figure 6.* Relationship between moose density estimated with cameras and with aerial surveys (solid line, shaded area is 90% confidence interval, $r^2$ = 0.84). The dashed line represents the 1:1 relationship. Error bars represent a 90% confidence interval in both the aerial and camera estimate.

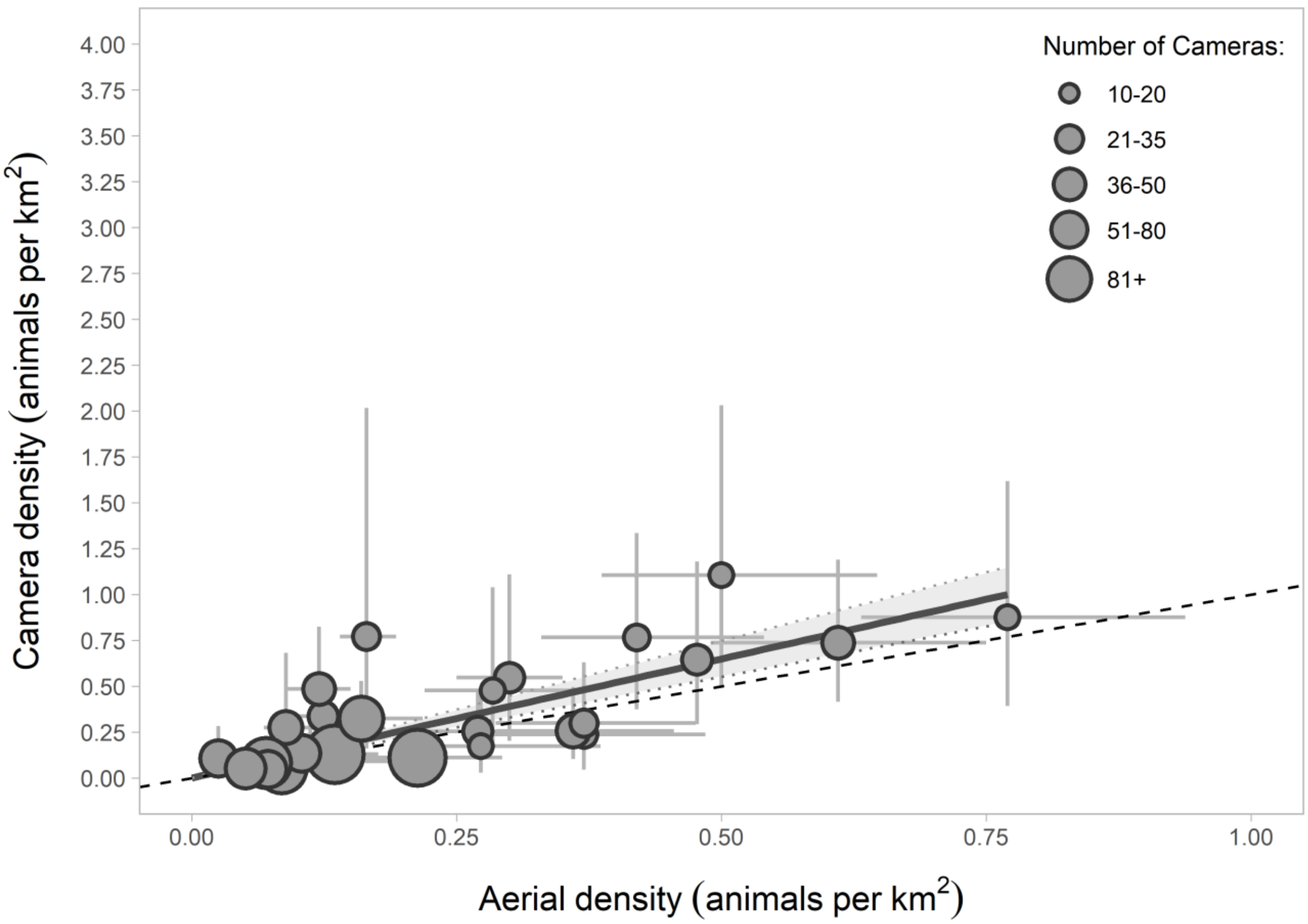

*Figure 7.* Relationship between moose density estimated with cameras and with aerial surveys (solid line, shaded area is the 90% confidence interval, $r^2 = 0.80$), with adjustment made to the camera estimates to account for predicted time investigating the camera and pole. The dotted line represents the 1:1 relationship. Error bars represent 90% confidence intervals in both the aerial and camera estimates.

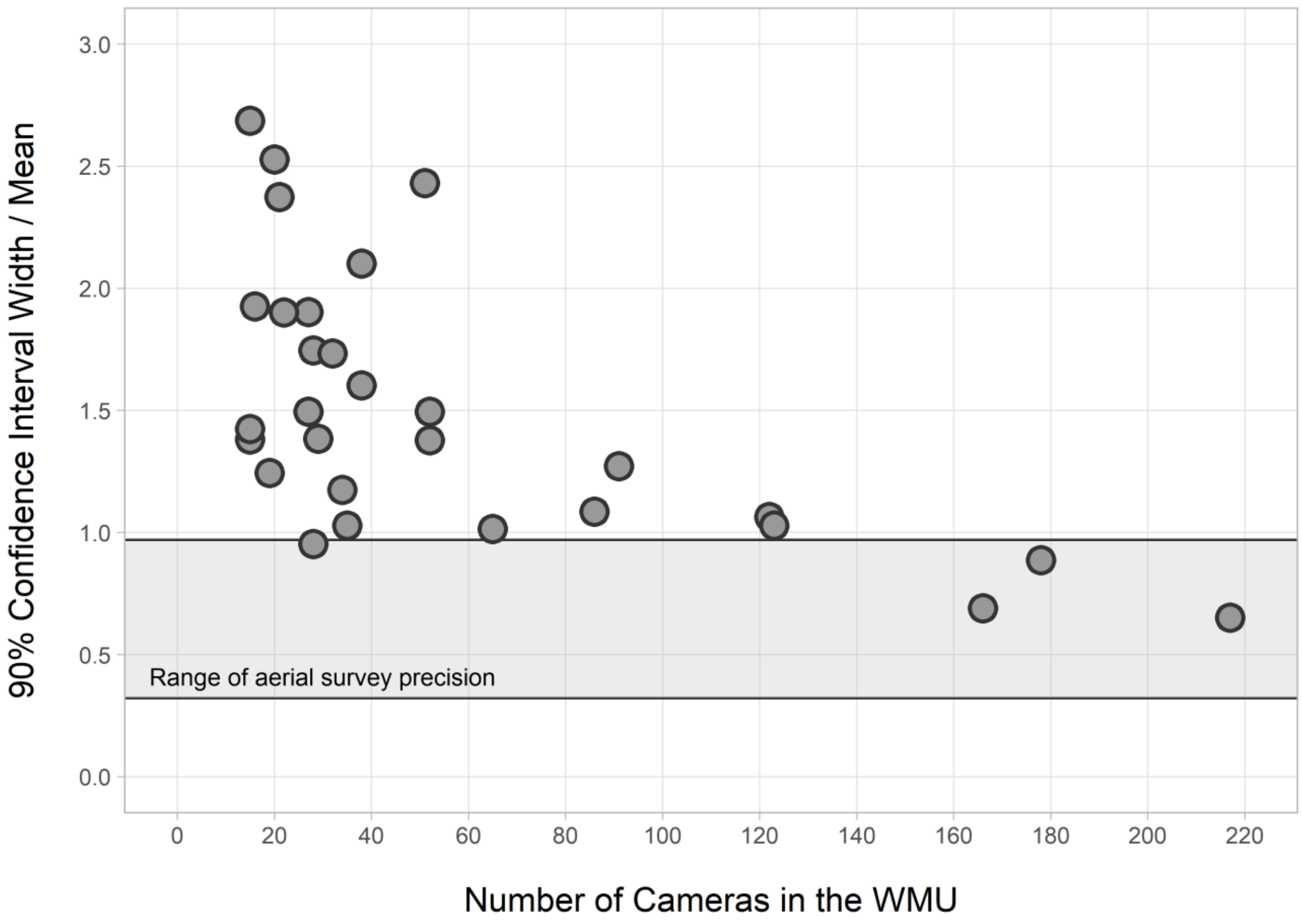

*Figure 8.* Relationship between the number of cameras in each WMU and the precision of the TIFC moose density estimate. The range of precision of aerial density estimates is shown by the shaded area. Precision is calculated as the width of the 90% confidence intervals (upper bound minus the lower bound) divided by the mean.

**Appendix S1. Effective Detection Distance**

Effective detection distances (EDD) are presented for each species group for different habitat types in the summer and winter seasons.

Table A1.1. Effective detection distance (EDD) in meters (m) for each species, or species group, by habitat type and season.

| **Species Group** | **Habitat Types** | **Summer** | **Winter** |
|---|---|---|---|
| Coyote and Fox [a] | Open Wetland | 10.39 | 10.35 |
| | Shrub | 7.01 | 6.99 |
| | Treed Wetland | 7.06 | 7.04 |
| | Deciduous | 7.07 | 7.05 |
| | Coniferous | 7.06 | 7.05 |
| | Human Footprint | 10.40 | 10.35 |
| | Water | 10.39 | 10.35 |
| | Grassland | 10.40 | 10.35 |
| Lynx [b] | Open Wetland | 7.29 | 7.79 |
| | Shrub | 6.12 | 6.39 |
| | Treed Wetland | 5.85 | 6.06 |
| | Deciduous | 5.84 | 6.05 |
| | Coniferous | 5.84 | 6.05 |
| | Human Footprint | 7.29 | 7.79 |
| | Water | 7.29 | 7.79 |
| | Grassland | 7.29 | 7.79 |
| Bear [c] | Open Wetland | 5.00 | 5.00 |
| | Shrub | 6.53 | 6.53 |
| | Treed Wetland | 5.78 | 5.78 |
| | Deciduous | 6.01 | 6.01 |
| | Coniferous | 5.58 | 5.58 |
| | Human Footprint | 7.15 | 7.15 |
| | Water | 5.00 | 5.00 |
| | Grassland | 11.83 | 11.83 |
| Bighorn Sheep [d] | Open Wetland | 5.80 | 5.80 |
| | Shrub | 5.80 | 5.80 |
| | Treed Wetland | 5.80 | 5.80 |
| | Deciduous | 5.80 | 5.80 |
| | Coniferous | 5.80 | 5.80 |
| | Human Footprint | 5.80 | 5.80 |
| | Water | 5.80 | 5.80 |
| | Grassland | 5.80 | 5.80 |
| Elk [e] | Open Wetland | 8.45 | 8.45 |
| | Shrub | 8.45 | 8.45 |
| | Treed Wetland | 8.45 | 8.45 |
| | Deciduous | 7.02 | 7.02 |
| | Coniferous | 7.02 | 7.02 |
| | Human Footprint | 10.21 | 10.21 |
| | Water | 8.45 | 8.45 |
| | Grassland | 8.45 | 8.45 |
| Mule Deer [f] | Open Wetland | 10.86 | 10.85 |
| | Shrub | 15.00 | 15.00 |
| | Treed Wetland | 6.97 | 6.96 |
| | Deciduous | 6.57 | 6.57 |
| | Coniferous | 7.50 | 7.50 |
| | Human Footprint | 7.67 | 7.67 |
| | Water | 11.46 | 11.45 |
| | Grassland | 7.56 | 7.56 |
| White-tailed Deer [g] | Open Wetland | 7.64 | 7.76 |
| | Shrub | 7.44 | 7.54 |
| | Treed Wetland | 8.17 | 8.31 |
| | Deciduous | 7.11 | 7.20 |
| | Coniferous | 7.42 | 7.53 |
| | Human Footprint | 8.45 | 8.59 |
| | Water | 6.93 | 7.01 |
| | Grassland | 7.86 | 7.98 |
| Pronghorn [h] | Open Wetland | 9.64 | 9.64 |
| | Shrub | 9.64 | 9.64 |
| | Treed Wetland | 9.64 | 9.64 |
| | Deciduous | 9.64 | 9.64 |
| | Coniferous | 9.64 | 9.64 |
| | Human Footprint | 9.64 | 9.64 |
| | Water | 9.64 | 9.64 |
| | Grassland | 9.64 | 9.64 |
| Small Species, Forest [i] | Open Wetland | 5.43 | 5.67 |
| | Shrub | 5.42 | 5.66 |
| | Treed Wetland | 5.41 | 5.64 |
| | Deciduous | 5.40 | 5.63 |
| | Coniferous | 5.40 | 5.63 |
| | Human Footprint | 5.46 | 5.71 |
| | Water | 5.43 | 5.67 |
| | Grassland | 5.43 | 5.67 |
| Small Species, Open Habitat [j] | Open Wetland | 6.08 | 6.09 |
| | Shrub | 6.08 | 6.08 |
| | Treed Wetland | 5.84 | 5.84 |
| | Deciduous | 5.83 | 5.83 |
| | Coniferous | 5.83 | 5.83 |
| | Human Footprint | 6.08 | 6.09 |
| | Water | 6.08 | 6.09 |
| | Grassland | 6.08 | 6.09 |
| Large Forest Carnivores [k] | Open Wetland | 6.38 | 6.16 |
| | Shrub | 6.37 | 6.15 |
| | Treed Wetland | 6.39 | 6.17 |
| | Deciduous | 6.39 | 6.17 |
| | Coniferous | 6.39 | 6.17 |
| | Human Footprint | 6.38 | 6.16 |
| | Water | 6.38 | 6.16 |
| | Grassland | 6.38 | 6.16 |
| Large Forest Ungulates [l] | Open Wetland | 6.48 | 6.48 |
| | Shrub | 6.49 | 6.49 |
| | Treed Wetland | 6.51 | 6.51 |
| | Deciduous | 6.51 | 6.51 |
| | Coniferous | 6.51 | 6.51 |
| | Human Footprint | 6.51 | 6.51 |
| | Water | 6.48 | 6.48 |
| | Grassland | 6.48 | 6.48 |

[a] Coyote (*Canis latrans*) and red foxe (*Vulpes vulpes*)
[b] Canada lynx (*Lynx canadensis*) and bobcat (*Lynx rufus*),
[c] Black (*Ursus americanus*) and grizzly bear (*Ursus arctos*)
[d] Bighorn sheep (*Ovis canadensis*), and mountain goat (*Oreamnos americanus*)
[e] Elk includes only elk (wapiti, *Cervus canadensis*),
[f] Mule Deer includes only mule deer (*Odocoileus hemionus*),
[g] White-tailed Deer' only includes white-tailed deer (*Odocoileus virgnianus*),
[h] Pronghorn includes only pronghorn (*Antilocapra americana*)
[i] Snowshoe hare (*Lepus americanus*), porcupine (*Erethizon dorsatum*), American red squirrel (*Tamiasciurus hudsonicus*), weasels and ermine (*Mustela* spp.), marten (*Martes americana*), fisher (*Pekania pennanti*), mink (*Neovison vison*), and northern flying squirrel (*Glaucomys sabrinus*)
[j] Raccoon (*Procyon lotor*), striped skunk (*Mephitis mephitis*), white-tailed jackrabbit (*Lepus townsendii),* muskrat (*Ondatra zibethicus*), Richardson's ground squirrel (*Urocitellus richardsonii*), badger (*Taxidea taxus*), hoary marmot (*Marmota caligata*), Columbian ground squirrel (*Urocitellus columbianus*), least chipmunk (*Neotamias minimus*), mice voles and allies (Family Muridae), groundhog (*Marmota monax*), beaver (*Castor canadensis*), golden-mantled ground squirrel (*Callospermophilus lateralis*), and river otter (*Lontra canadensis*)
[k] Cougar (*Puma concolor*), gray wolf (*Canis lupus*), and wolverine (*Gulo gulo*)
[l] Moose (*Alces alces*), bison (*Bison bison*), and woodland caribou (*Rangifer tarandus*).

## Appendix S2. Probability of Leaving Field-of-view

Results for the probability of an animal leaving the field-of-view for intervals of 20-120 seconds between images are presented below for six species groups. To account for small sample sizes, we grouped together similar species with lower sample size in the modeling procedure. The following groups were used:

- ‘Bears’ includes black (*Ursus americanus*) and grizzly bear (*Ursus arctos*).
- ‘Canids and Cougars’ includes cougar (*Puma concolor*), coyote (*Canis latrans*), gray wolf (*Canis lupus*), and red fox (*Vulpes vulpes*).
- ‘Moose’ includes only moose (*Alces alces*).
- ‘Other Ungulates’ includes bighorn sheep (*Ovis canadensis*), bison (*Bison bison*), white-tailed deer (*Odocoileus virgnianus*), mule deer (*Odocoileus hemionus*), elk (wapiti, *Cervus canadensis*), mountain goat (*Oreamnos americanus*), woodland caribou (*Rangifer tarandus*), and pronghorn (*Antilocapra americana*).
- ‘Small Carnivores’ includes badger (*Taxidea taxus*), canada lynx (*Lynx canadensis*), fisher (*Pekania pennanti*), marten (*Martes americana*), mink (*Neovison vison*), raccoon (*Procyon lotor*), and wolverine (*Gulo gulo*).
- ‘Small Mammals’ includes groundhog (*Marmota monax*), chipmunk (*Tamias* spp.), porcupine (*Erethizon dorsatum*), snowshoe hare (*Lepus americanus*), skunk (*Mephitis mephitis*), and white-tailed jack rabbit (*Lepus townsendii).*

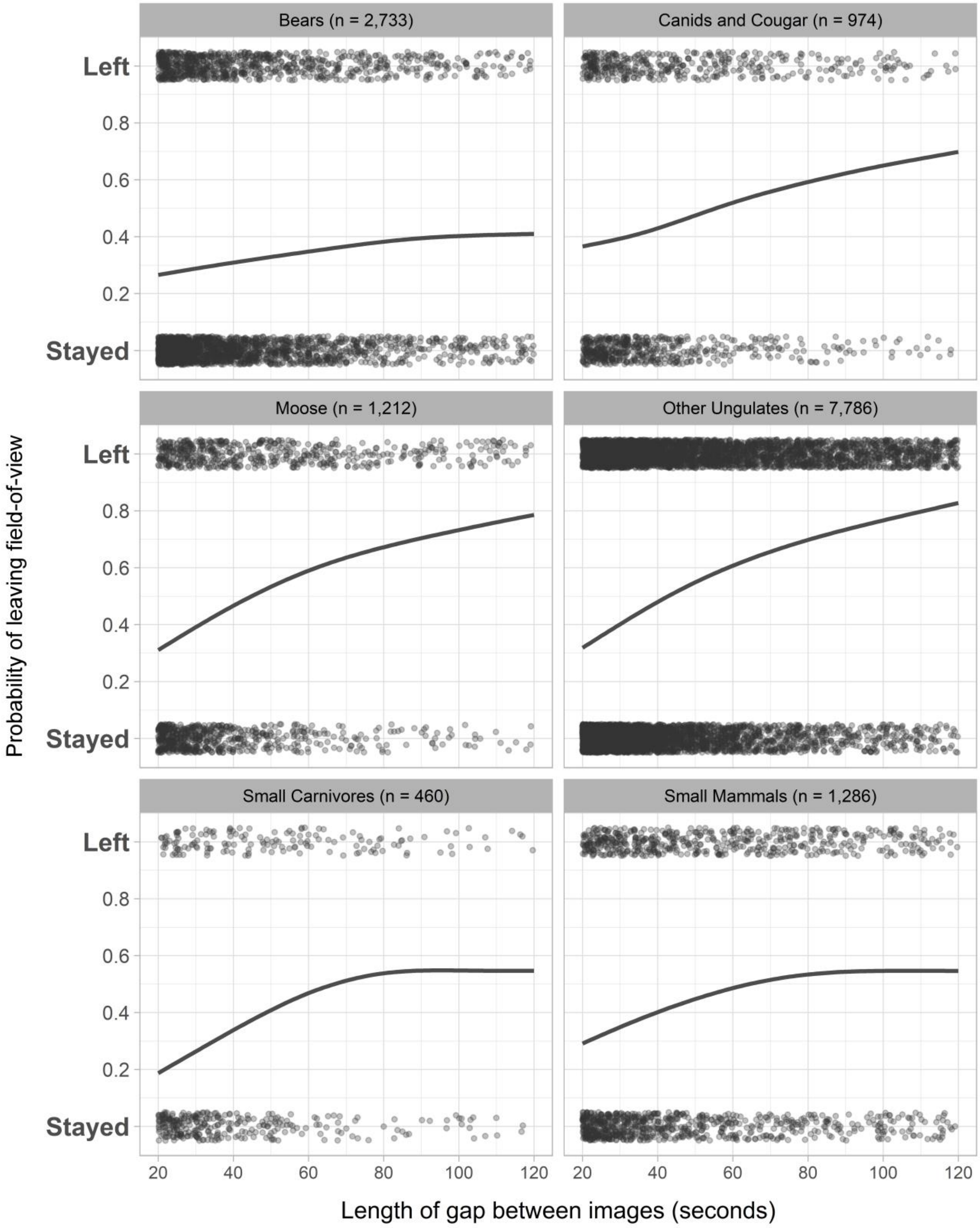

Figure A2. Probability an animal left the camera field-of-view as a function of interval between consecutive images, by species group.

**Appendix S3. Micro-Habitat at Camera Sites and Density**

This appendix presents densities of seven species in different micro-habitats in two broad stand types, as a test of violating the assumption of representative sampling locations.

Table A3.1. Estimated density (animals per km$^2$) and corresponding 90% confidence intervals (CI) of each species in treed, low-productivity open and productive open micro-habitat in deciduous and coniferous forests.

| **Species** | **Micro-habitat** | **Density (90% CI)** | |
|---|---|---|---|
| | | **Deciduous** | **Coniferous** |
| Moose | Treed | 0.58 (0.28, 0.95) | 0.21 (0.09, 0.36) |
| | Open (Low) | 0.28 (0.07, 0.59) | 0.16 (0.08, 0.29) |
| | Open (Productive) | 0.94 (0.45, 1.55) | 0.76 (0.35, 1.26) |
| Black Bear | Treed | 0.67 (0.31, 1.15) | 1.09 (0.48, 1.84) |
| | Open (Low) | 1.09 (0.55, 1.77) | 0.59 (0.34, 0.92) |
| | Open (Productive) | 1.27 (0.60, 2.22) | 0.72 (0.36, 1.18) |
| Coyote | Treed | 0.02 (0.01, 0.03) | 0.00 (0.00, 0.01) |
| | Open (Low) | 0.01 (0.00, 0.02) | 0.01 (0.00, 0.02) |
| | Open (Productive) | 0.04 (0.02, 0.05) | 0.02 (0.00, 0.04) |
| Fisher | Treed | 0.01 (0.00, 0.02) | 0.01 (0.00, 0.03) |
| | Open (Low) | 0.01 (0.00, 0.03) | 0.01 (0.00, 0.02) |
| | Open (Productive) | 0.04 (0.01, 0.10) | 0.00 (0.00, 0.01) |
| Gray Wolf | Treed | 0.03 (0.00, 0.08) | 0.01 (0.00, 0.04) |
| | Open (Low) | 0.09 (0.00, 0.18) | 0.02 (0.00, 0.01) |
| | Open (Productive) | 0.01 (0.00, 0.03) | 0.01 (0.00, 0.05) |
| Snowshoe Hare | Treed | 0.42 (0.20, 0.69) | 0.30 (0.16, 0.49) |
| | Open (Low) | 0.39 (0.10, 0.80) | 0.78 (0.40, 1.20) |
| | Open (Productive) | 0.33 (0.17, 0.51) | 0.41 (0.18, 0.71) |
| White-tailed Deer | Treed | 1.09 (0.59, 1.69) | 0.32 (0.17, 0.53) |
| | Open (Low) | 0.88 (0.21, 1.70) | 0.48 (0.23, 0.83) |
| | Open (Productive) | 1.77 (1.06, 2.59) | 0.81 (0.47, 1.69) |

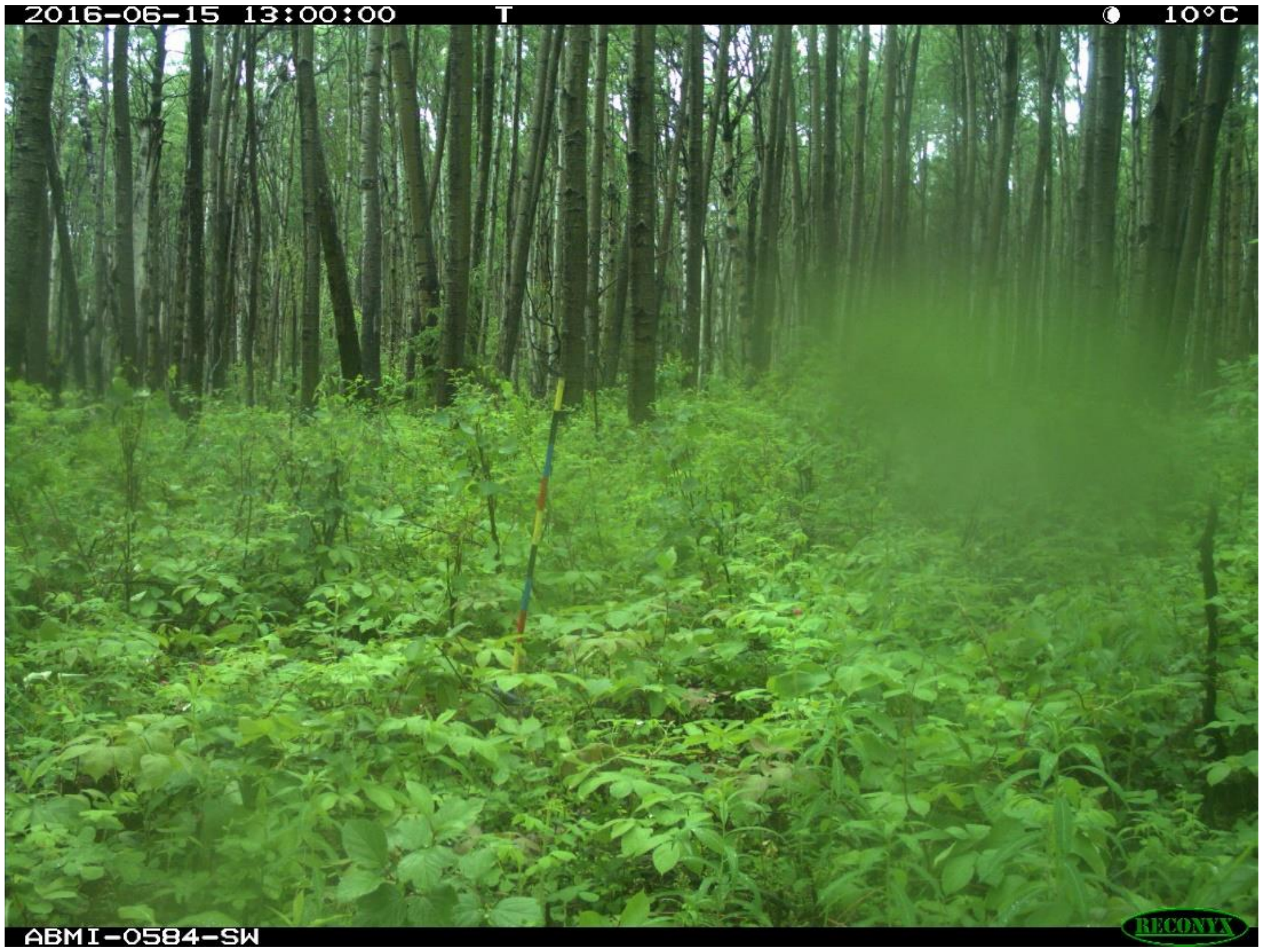
2016-06-15 13:00:00
T
10°C
ABMI-0584-SW
RECONYX

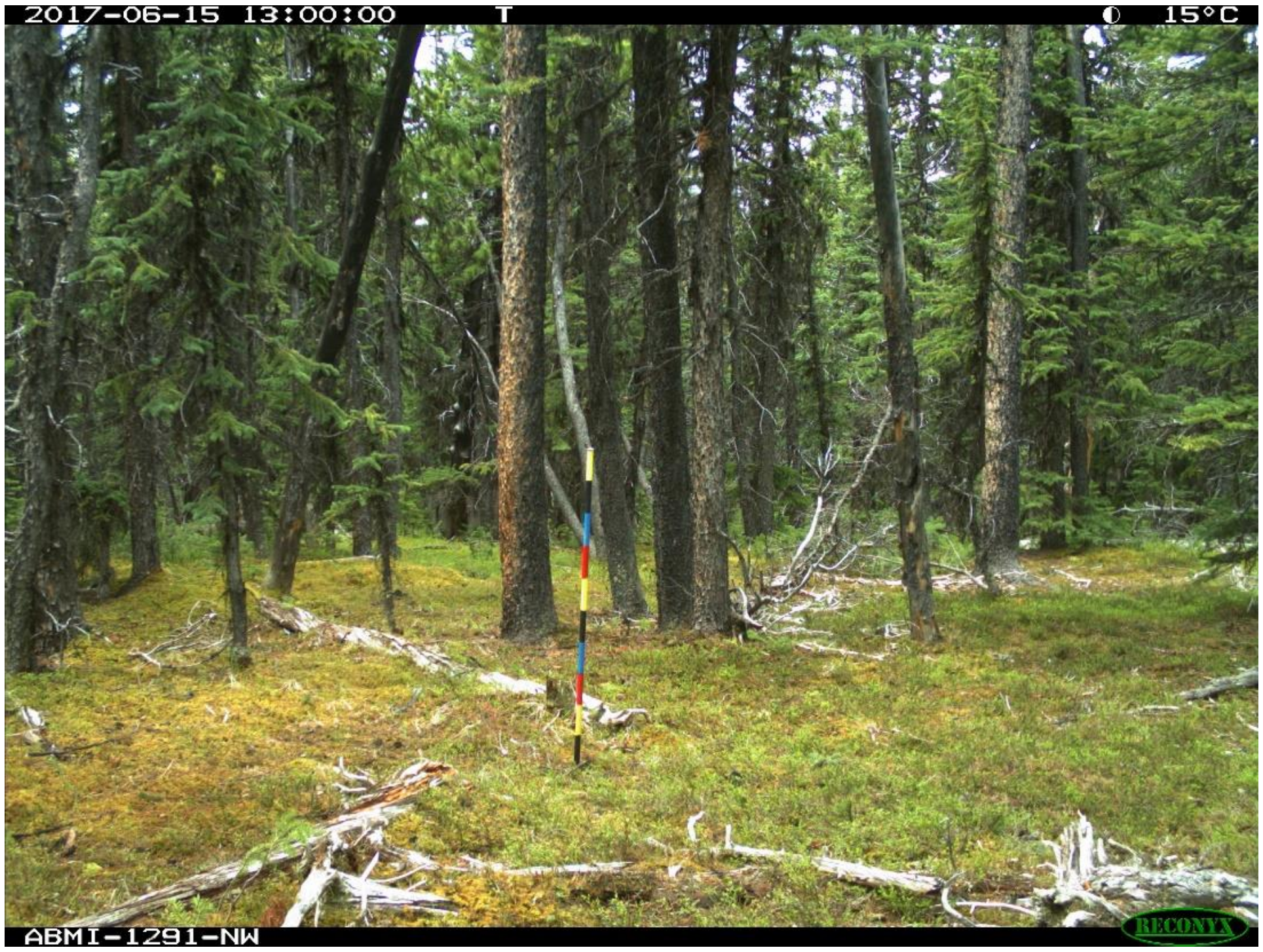
2017-06-15 13:00:00
T
15°C
ABMI-1291-NW
RECONYX

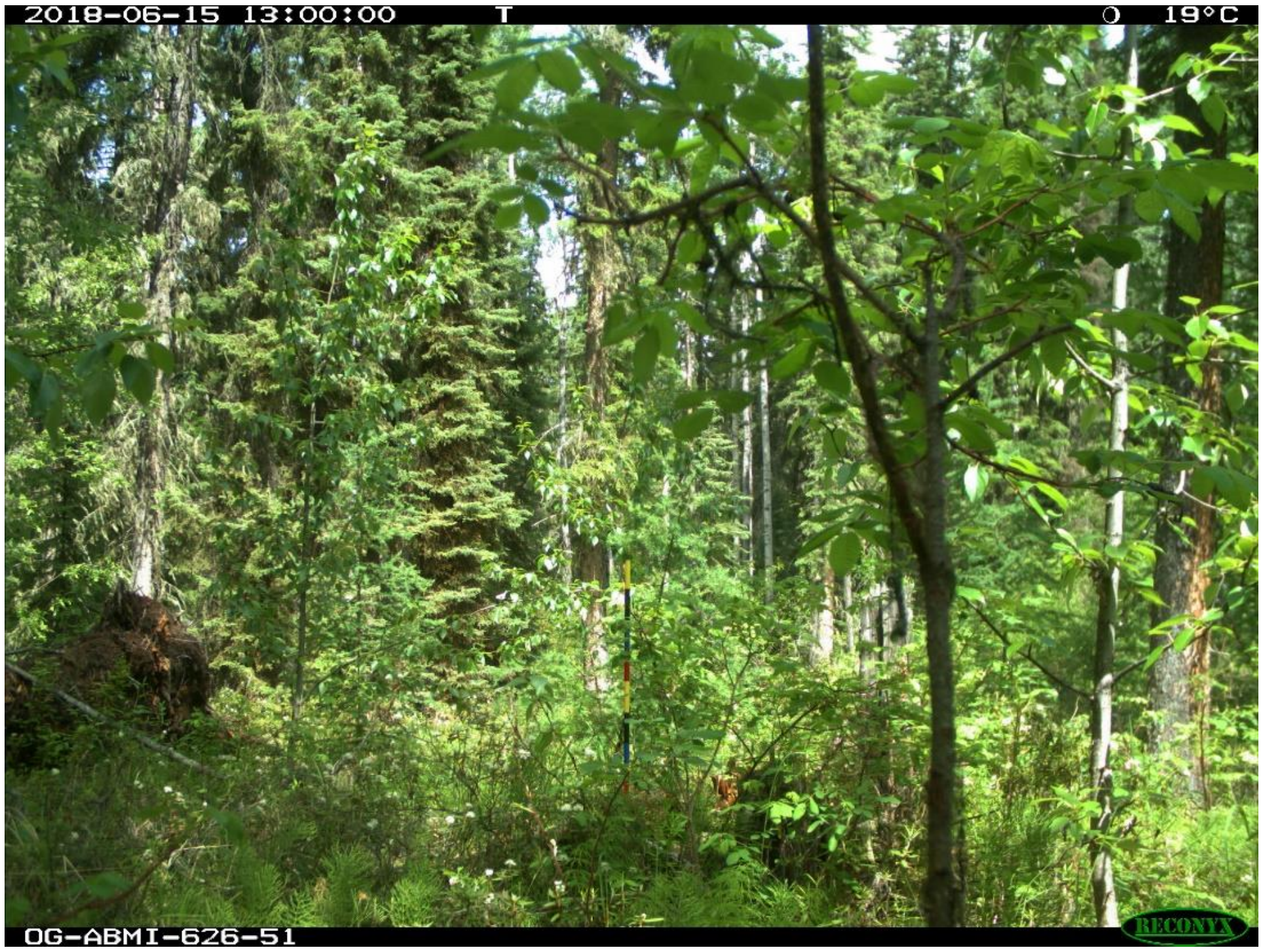

Figure A3. Examples of open productive (top), open low-productivity (middle), and treed (bottom) micro-habitat camera sites.

**Appendix S4. Camera Investigation Behaviors**

This appendix presents the proportion of time species spend investigating the camera or 5m pole, and the time investigating + associated behaviors, in several habitat types, as a test of the assumption that movement is not affected by the camera.

Table A4.1. The estimated proportion of time, and 90% confidence intervals (CI), spent by each species investigating the camera, or investigating the camera and associated behavior, by habitat type.

| **Species** | **Habitat Type** | **Proportion Time (90% CI)** | |
|---|---|---|---|
| | | **Investigating Time** | **Investigating + Associated** |
| Moose | All | 0.51 (0.46, 0.55) | 0.67 (0.62, 0.71) |
| | Coniferous | 0.48 (0.35, 0.62) | 0.58 (0.45, 0.71) |
| | Deciduous | 0.34 (0.25, 0.43) | 0.47 (0.38, 0.56) |
| | Grassland | 0.71 (0.51, 0.85) | 0.83 (0.64, 0.93) |
| | Human Footprint | 0.51 (0.35, 0.66) | 0.56 (0.40, 0.71) |
| | Shrub | 0.57 (0.45, 0.68) | 0.86 (0.75, 0.92) |
| | Open Wetland | 0.59 (0.40, 0.76) | 0.79 (0.59, 0.91) |
| | Treed Wetland | 0.63 (0.51, 0.74) | 0.81 (0.70, 0.89) |
| Black Bear | All | 0.59 (0.54, 0.63) | 0.67 (0.62, 0.72) |
| | Coniferous | 0.48 (0.40, 0.57) | 0.57 (0.49, 0.66) |
| | Deciduous | 0.63 (0.55, 0.71) | 0.70 (0.62, 0.77) |
| | Grassland | - | - |
| | Human Footprint | 0.46 (0.24, 0.70) | 0.46 (0.24, 0.70) |
| | Shrub | 0.52 (0.35, 0.69) | 0.74 (0.55, 0.87) |
| | Open Wetland | 0.77 (0.57, 0.89) | 0.83 (0.64, 0.93) |
| | Treed Wetland | 0.70 (0.56, 0.81) | 0.76 (0.63, 0.86) |
| White-tailed Deer | All | 0.12 (0.09, 0.15) | 0.22 (0.18, 0.26) |
| | Coniferous | 0.05 (0.01, 0.27) | 0.06 (0.01, 0.27) |
| | Deciduous | 0.05 (0.02, 0.12) | 0.24 (0.17, 0.34) |
| | Grassland | 0.15 (0.08, 0.27) | 0.24 (0.14, 0.36) |
| | Human Footprint | 0.17 (0.11, 0.25) | 0.23 (0.16, 0.32) |
| | Shrub | 0.10 (0.03, 0.27) | 0.25 (0.13, 0.43) |
| | Open Wetland | 0.06 (0.01, 0.46) | 0.10 (0.01, 0.46) |
| | Treed Wetland | 0.12 (0.06, 0.24) | 0.20 (0.11, 0.33) |
| Mule Deer | All | 0.11 (0.08, 0.14) | 0.16 (0.13, 0.21) |
| | Coniferous | 0.24 (0.08, 0.55) | 0.29 (0.11, 0.59) |
| | Deciduous | 0.10 (0.01, 0.46) | 0.13 (0.02, 0.47) |
| | Grassland | 0.08 (0.04, 0.14) | 0.11 (0.07, 0.17) |
| | Human Footprint | 0.11 (0.07, 0.17) | 0.18 (0.13, 0.25) |
| | Shrub | 0.14 (0.05, 0.35) | 0.18 (0.07, 0.40) |
| | Open Wetland | 0.16 (0.07, 0.33) | 0.26 (0.14, 0.44) |
| | Treed Wetland | - | - |
| Gray Wolf | All | 0.17 (0.13, 0.23) | 0.31 (0.25, 0.37) |
| | Coniferous | 0.39 (0.26, 0.53) | 0.45 (0.32, 0.60) |
| | Deciduous | 0.10 (0.04, 0.24) | 0.37 (0.24, 0.53) |
| | Grassland | - | - |
| | Human Footprint | 0.05 (0.01, 0.31) | 0.15 (0.04, 0.39) |
| | Shrub | 0.26 (0.16, 0.41) | 0.30 (0.18, 0.45) |
| | Open Wetland | 0.11 (0.03, 0.31) | 0.32 (0.17, 0.53) |
| | Treed Wetland | 0.04 (0.01, 0.13) | 0.20 (0.11, 0.32) |

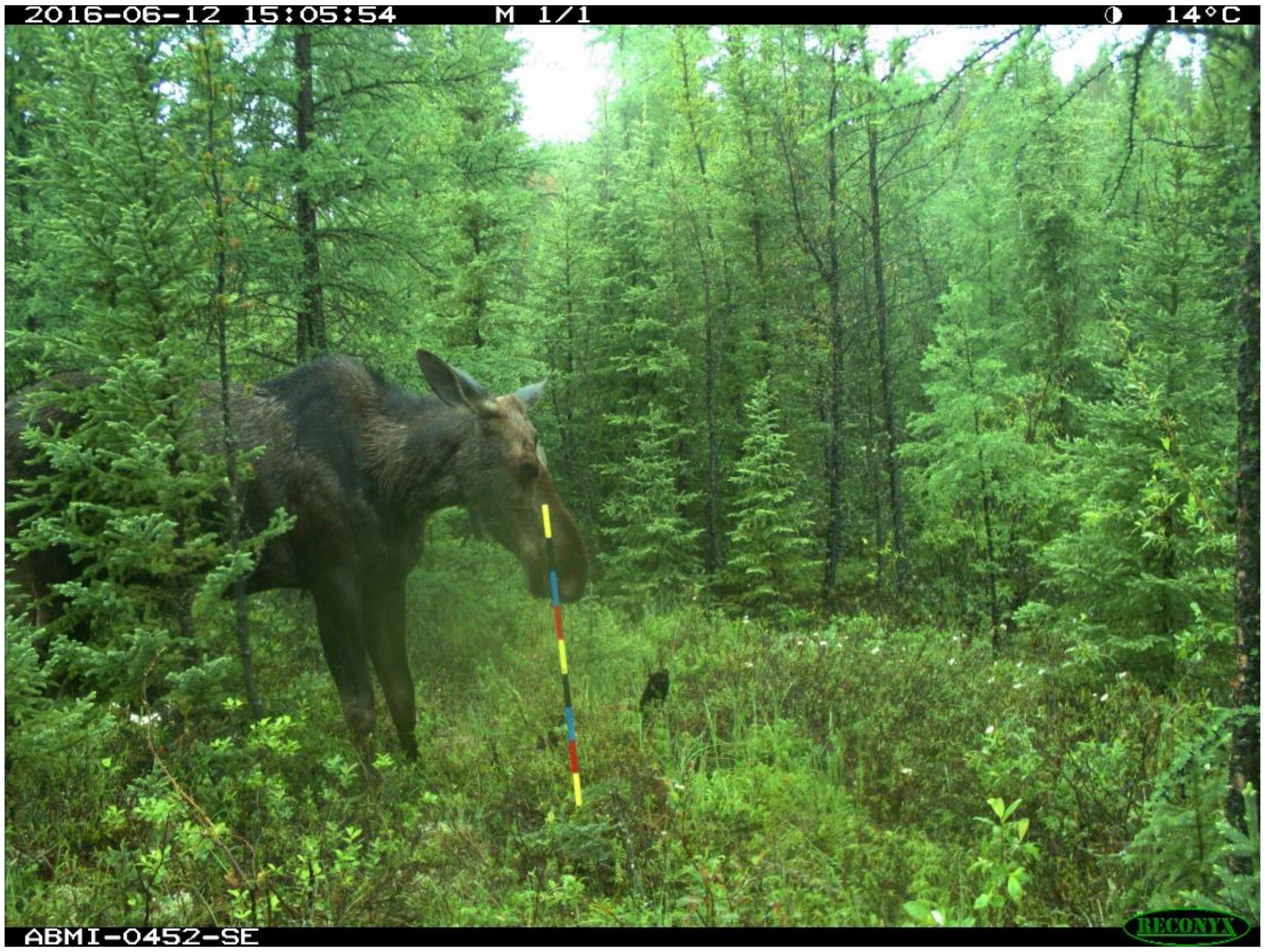

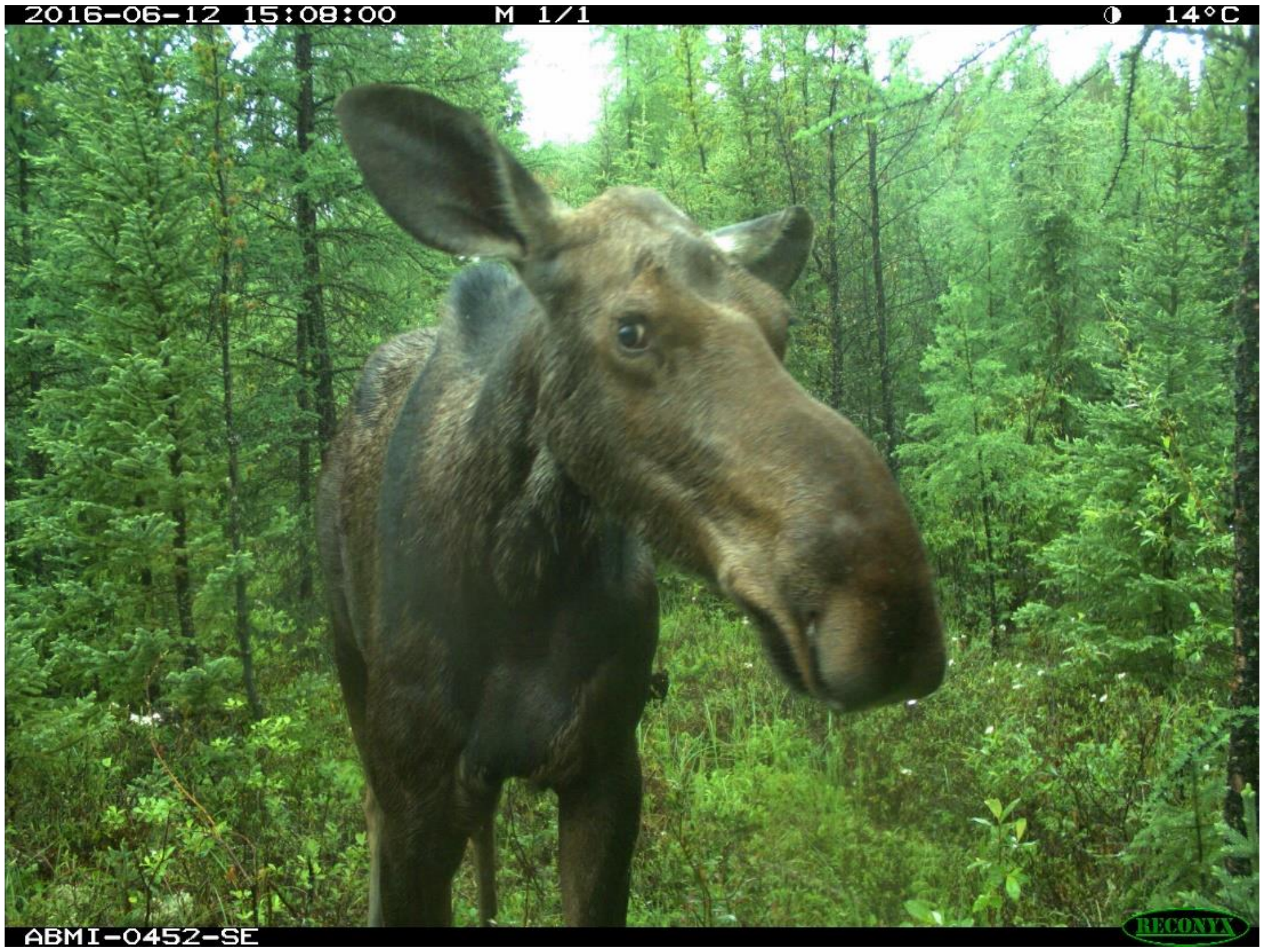
2016-06-12 15:08:00
M 1/1
14°C
ABMI-0452-SE
RECONYX

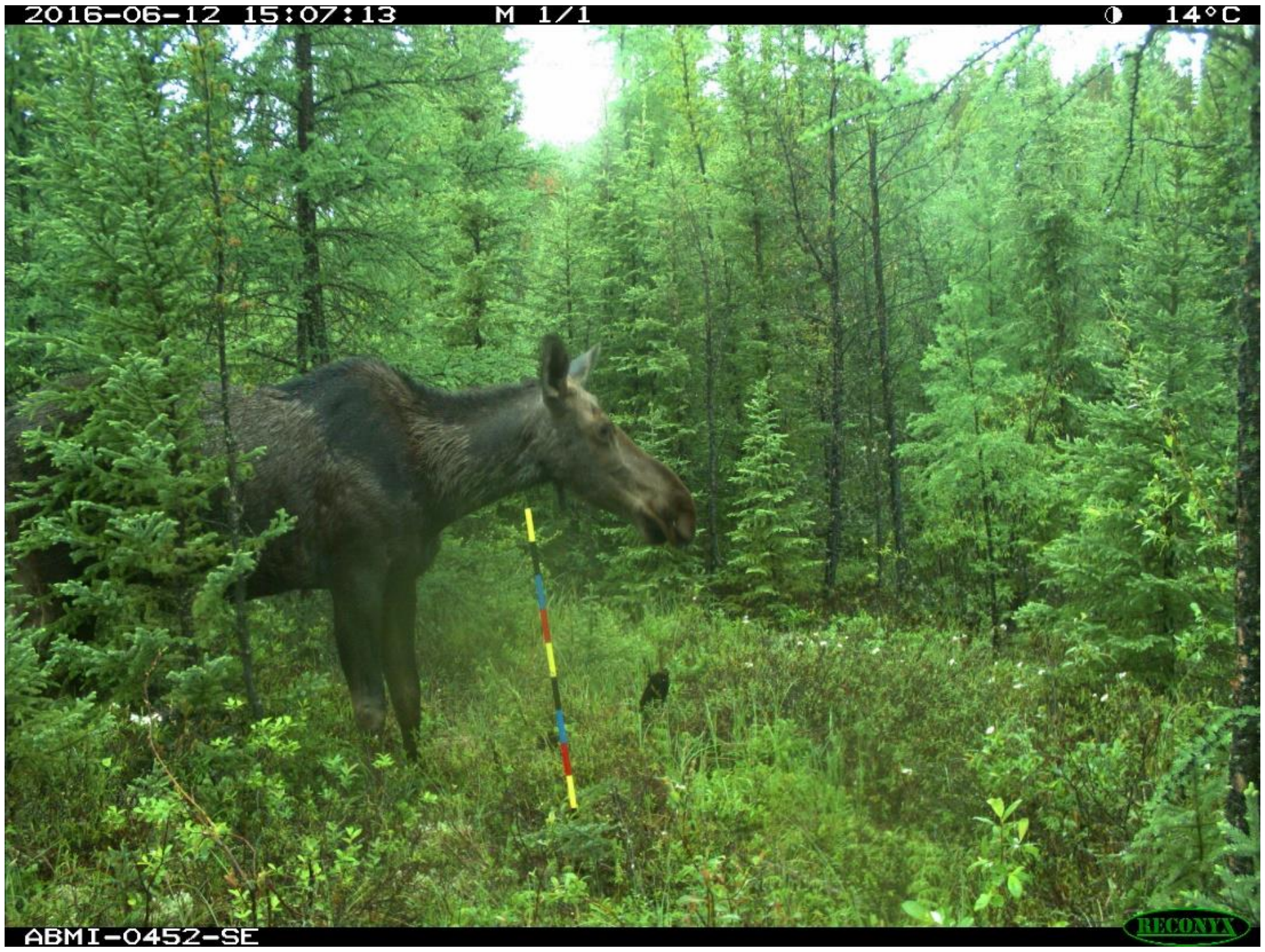
2016-06-12 15:07:13
M 1/1
14°C
ABMI-0452-SE
RECONYX

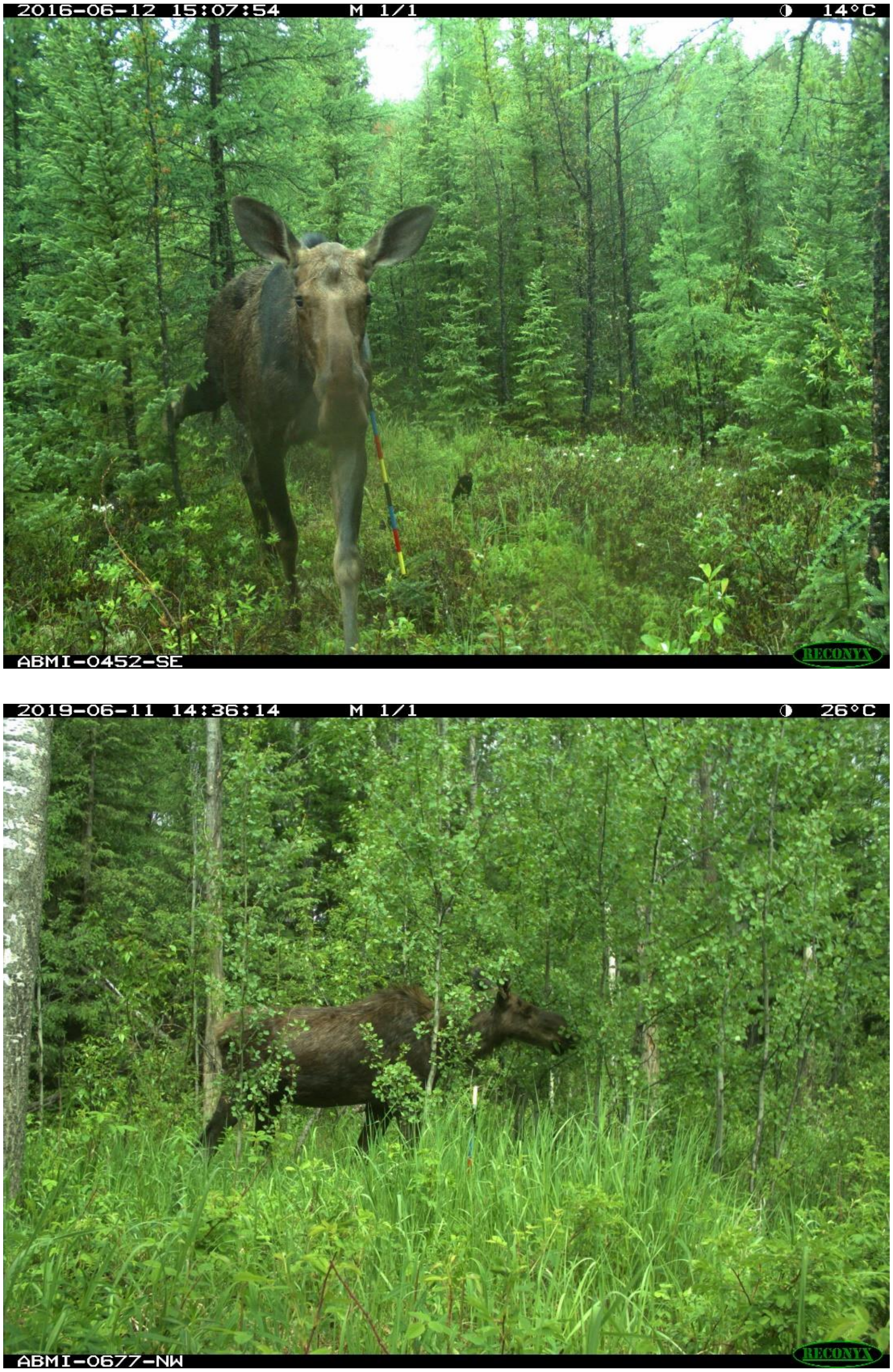

Figure A4.1. From top to bottom, example of a moose investigating the 5m pole (behavior 1), investigating the camera (behaviour 1), lingering after having investigated the pole (behavior 2), directly traveling to investigate the camera (behavior 2), and natural activity (behavior 3).

**Appendix S5. Effects of Lures**

Lure results (ratio of lured:unlured) are presented for each species, for: occurrence (presence/absence), density given occurrence (density at deployment where the species occurred), and total density.

Table A5.1. Ratios of lured:unlured for occurrence, density given occurrence, and total estimated density for each species.

| **Species** | **Occurrence** | **Density given Occurrence** | **Total Density** |
|---|---:|---:|---:|
| Black Bear | 1.34 (1.24-1.45) | 2.80 (2.23-3.57) | 3.75 (2.96-4.83) |
| Lynx | 1.43 (1.27-1.62) | 1.76 (1.32-2.31) | 2.52 (1.86-3.37) |
| Coyote | 1.38 (1.28-1.49) | 3.38 (2.79-4.14) | 4.67 (3.84-5.71) |
| Elk | 1.09 (0.92-1.29) | 1.43 (0.97-1.94) | 1.55 (1.07-2.08) |
| Fisher | 4.60 (3.45-6.55) | 2.45 (1.73-3.79) | 11.25 (7.25-20.39) |
| Gray Wolf | 1.55 (1.26-1.94) | 3.66 (2.23-6.07) | 5.68 (3.31-9.9) |
| Marten | 2.38 (1.97-2.93) | 1.38 (0.92-2.19) | 3.29 (2.19-5.36) |
| Moose | 1.07 (0.99-1.16) | 1.17 (0.92-1.50) | 1.26 (0.98-1.62) |
| Mule Deer | 1.03 (0.93-1.13) | 1.97 (1.47-2.57) | 2.03 (1.51-2.62) |
| Red Fox | 2.16 (1.78-2.67) | 3.68 (2.42-5.77) | 7.94 (5.14-12.88) |
| Red Squirrel | 1.47 (1.20-1.80) | 0.59 (0.31-1.24) | 0.87 (0.44-1.19) |
| Snowshoe Hare | 1.00 (0.93-1.09) | 1.06 (0.84-1.36) | 1.07 (0.84-1.37) |
| White-tailed Deer | 1.03 (0.99-1.08) | 0.87 (0.71-1.06) | 0.90 (0.74-1.09) |

## Appendix S6. Summary of Moose Density Estimates

Moose aerial density estimates used in the Wildlife Management Unit (WMU) application, including confidence intervals, were obtained from *https://www.alberta.ca/aerial-wildlife-survey-reports.aspx* (Accessed July 21, 2021).

Table A6.1. Moose density estimates for each Wildlife Management Unit (WMU) from cameras and aerial surveys, and the number of cameras used to estimate density.

| WMU Code | Number of Cameras | Density (Camera) | Density (Aerial) |
|---|---|---|---|
| 00256 | 27 | 0.22 (0.06, 0.47) | 0.11 (0.08, 0.16) |
| 00258 | 21 | 1.23 (0.27, 3.20) | 0.17 (0.14, 0.19) |
| 00349 | 20 | 0.44 (0.08, 1.20) | 0.37 (0.28, 0.49) |
| 00356 | 35 | 1.38 (0.77, 2.20) | 0.61 (0.49, 0.75) |
| 00357 | 38 | 0.59 (0.23, 1.17) | 0.36 (0.29, 0.46) |
| 00358 | 15 | 1.60 (0.71, 2.92) | 0.77 (0.63, 0.94) |
| 00502 | 27 | 0.50 (0.21, 0.95) | 0.27 (0.21, 0.34) |
| 00503 | 28 | 1.26 (0.44, 2.60) | 0.30 (0.25, 0.35) |
| 00504 | 15 | 1.79 (0.77, 3.30) | 0.50 (0.39, 0.65) |
| 00509 | 22 | 0.62 (0.19, 1.36) | 0.37 (0.29, 0.48) |
| 00511 | 28 | 0.61 (0.36, 0.94) | 0.12 (0.09, 0.18) |
| 00512 | 166 | 0.24 (0.17, 0.33) | 0.21 (0.16, 0.29) |
| 00514 | 15 | 0.31 (0.05, 0.88) | 0.27 (0.19, 0.39) |
| 00515 | 32 | 0.23 (0.08, 0.49) | 0.14 (0.10, 0.19) |
| 00516 | 91 | 0.20 (0.10, 0.35) | 0.08 (0.06, 0.10) |
| 00517 | 122 | 0.12 (0.07, 0.20) | 0.09 (0.07, 0.11) |
| 00519 | 178 | 0.29 (0.18, 0.44) | 0.14 (0.10, 0.18) |
| 00520 | 86 | 0.54 (0.30, 0.89) | 0.16 (0.12, 0.22) |
| 00523 | 19 | 1.57 (0.76, 2.70) | 0.42 (0.33, 0.54) |
| 00526 | 29 | 1.23 (0.55, 2.25) | 0.48 (*NA*, *NA*) |
| 00527 | 16 | 0.94 (0.29, 2.10) | 0.28 (0.22, 0.37) |
| 00528 | 52 | 0.25 (0.11, 0.46) | 0.10 (0.08, 0.13) |
| 00529 | 38 | 0.49 (0.13, 1.15) | 0.09 (0.07, 0.12) |
| 00530 | 217 | 0.39 (0.27, 0.53) | 0.12 (0.10, 0.14) |
| 00531 | 123 | 0.17 (0.09, 0.30) | 0.07 (0.05, 0.09) |
| 00540 | 51 | 0.24 (0.05, 0.64) | 0.03 (0.02, 0.04) |
| 00542 | 52 | 0.08 (0.03, 0.16) | 0.07 (0.06, 0.09) |
| 00544 | 34 | 0.81 (0.41, 1.36) | 0.12 (0.09, 0.15) |
| 00726 | 65 | 0.09 (0.05, 0.16) | 0.05 (0.04, 0.08) |